\documentclass[useAMS,usenatbib]{mn2e}
\usepackage[spanish,activeacute,english]{babel}
\usepackage[latin1]{inputenc}
\usepackage[dvips]{graphicx,color}
\usepackage{epsfig}
\usepackage{latexsym}
\usepackage{amssymb}
\usepackage{amsmath}
\usepackage{verbatim}
\usepackage{natbib}
\usepackage{caption}
\usepackage[T1]{fontenc} % Computer Modern (CM) fonts
\usepackage{ae,aecompl} % and: dvips -Pcmz -Pamz macros_aaa.dvi

\IfFileExists{.pdfmode}{%
  \def\UseColor{true}
  \usepackage{ae,aecompl} % and: dvips -Pcmz -Pamz *.dvi
  }{
  \def\UseColor{false}
  }

\usepackage[pdftitle={Stars on the run: escaping from stellar clusters},%
pdfauthor={Guido R. I. Moyano Loyola},%
pdfsubject={},%
pdfkeywords={},%
colorlinks=\UseColor,%
dvips%
]{hyperref}

\title[Stars on the run: escaping from stellar clusters]
  {Stars on the run: escaping from stellar clusters}
\author[G. R. I. Moyano Loyola, J. R. Hurley]
  {Guido R. I.~Moyano Loyola,$^1$\thanks{Email: gmoyano@astro.swin.edu.au}
  Jarrod R.~Hurley$^1$\thanks{Email: jhurley@astro.swin.edu.au} \\
  $^1$Centre for Astrophysics and Supercomputing, Swinburne University of Technology, PO Box 218, Australia}

\date{Accepted 2013 June 26.  Received 2013 June 23; in original form 2013 April 15}

\pagerange{\pageref{firstpage}--\pageref{lastpage}} \pubyear{2013}

\def\LaTeX{L\kern-.36em\raise.3ex\hbox{a}\kern-.15em
    T\kern-.1667em\lower.7ex\hbox{E}\kern-.125emX}

\begin{document}

\label{firstpage}

\maketitle

\begin{abstract}

A significant proportion of Milky Way stars are born in stellar clusters, which dissolve over time so that the members become part of the disc and halo populations of the Galaxy. In the present work we will assume that these young stellar clusters live mainly within the disc of the Galaxy and that they can have primordial binary percentages ranging from 0\% to as high as 70\%. We have evolved models of such clusters to an age of 4 Gyr through {\it{N}}-body simulations,
paying attention to the stars and binaries that escape in the
process. We have quantified the contribution of these escaping stars to
the Galaxy population by analysing their escape velocity and
evolutionary stage at the moment of escape.
In this way we could analyse the mechanisms that produced these
escapers, whether evaporation through weak two-body encounters,
energetic close encounters or stellar evolution events, e.g. supernovae. In our models we found that the percentage of primordial binaries
in a star cluster does not produce significant variations in the
velocities of the stars that escape in the velocity range of
0-20 km s$^{-1}$.
However, in the high-velocity 20-100 km s$^{-1}$ range  the number of
escapers increased markedly as the primordial binary percentage
increased. We could also infer that dissolving stellar clusters such as those
that we have modelled can populate the Galactic halo with giant 
stars for which the progenitors were stars of up to 2.4 M$_{\odot}$.
Furthermore, choices made for the velocity kicks of remnants do influence the production of hyper-velocity stars -- and to a lesser
extent stars in the high-velocity range -- but once again the
difference for the 99\% of stars in the 0-20 km s$^{-1}$ range is not
significant.

\end{abstract}

\begin{keywords}
methods: N-body simulations\ -- (Galaxy:) open clusters and associations: general < The Galaxy
\end{keywords}

\section{Introduction}

The pioneering work of \citet{vonH,vonH2} on the infant computation of {\it{N}}-body simulations with the first computers opened a new world of po\-ssi\-bi\-li\-ties in the study of stellar systems. This new world grew as the computational
 techniques improved over time (\citealt{Aarseth1}), the dynamical theories were refined (e.g. the classification of collisional versus collisionless models and related approximations) and faster computers were built.

As time went by numerical codes evolved to take into account a mass spectrum (e.g. \citealt{Salpeter}; \citealt{Kroupa}) for the {\it{N}}-bodies and the evolution of a system inside a given background potential (e.g. the evolution of a cluster in the potential generated by its host galaxy, \citealt{Hayli}). However, a crucial 
piece was still missing, stars intrinsically evolve on a timescale comparable to or smaller than the dynamical timescales. Indeed, the different timescales involved are a
 key aspect in the study of numerical simulations of stellar systems (\citealt{Meylan}): the dynamical timescale of the system as a whole, dynamical timescales of 
multiples systems such as binaries, triples or quartets and the intrinsic nuclear timescale of the stars, to name a few.

In the early 2000's when treatments for single and binary stellar evolution (\citealt{Hurley1}; \citealt{Hurley2}) were added into one of the most famous collisional {\it{N}}-body codes, NBODY6 
\citep{Aarseth3}, a sophisticated code incorporating all of the relevant timescales was available. This implementation built on earlier efforts to include a simplified stellar evolution prescription (e.g. \citealt{CW}; \citealt{EFT}) and binary evolution algorithm (e.g. \citealt{TAPE}) in dynamical simulations. It enabled the production of realistic mo\-dels of open clusters with {\it{N}} $\sim$ 30\,000 (e.g. \citealt{Hurley3}) and helped our understanding of star cluster evolution.

The identification of stars within star forming regions, and particularly those that are members of young clusters has been improved in recent years, especially with the use of high-resolution infrared instruments (e.g. {\it{Spitzer}}). It is commonly assumed that most, if not all, stars are born in stellar clusters or loose associations of some sort \citep{Lada}, but recent studies have shed new light on the fraction of stars that are actually born in a given star forming region and remain bound to form a stellar cluster (\citealt{Bressert}; \citealt{Kruijssen}).
In the event that stars do remain gravitationally bound to form a stellar cluster (i.e. open cluster), with the  outcome of being  dynamically and even chemically affected by their siblings, the bulk of these clusters will dissolve over time by two-body processes that will cause stars to reach their escape velocities and by the tidal stripping imposed by the external potential of the Galaxy in which these clusters are embedded. Quantifying the contribution of these escaping stars to the
 disc and halo populations is a key factor to be able to trace back the stellar po\-pu\-la\-tions that were the first buil\-ding blocks of our Galaxy \citep{Freeman} and in 
this way reconstruct the assembly history of it.

In the present work we will assume that these young stellar clusters live mainly within the disc of the Galaxy. The presence of the tidal field imposed by the Galaxy
 (\citealt{Holger1} and references therein) along with the stellar evolution that each star will undergo and the dynamical interaction between the stars, e.g. two-body interactions, will determine the evolution of these clusters. 

The outcome of the processes mentioned is that stars escape from stellar clusters (\citealt{Leonard1}; \citealt{Leonard2}; \citealt{Leonard3}). The study of the escaping time-scale of these stars is of great importance \citep{Fukushige2} as is their ve\-lo\-ci\-ty (\citealt{Kollmeier}; \citealt{Silva}; \citealt{Gvaramadze}; \citealt{Li}) and even population type \citep{Royer}, particularly
 the study of the OB stars (\citealt{PortegiesZwart}; \citealt{Berger}; \citealt{Hoogerwerf}) because these type of stars are easier to detect in kinematical and
 proper motion stu\-dies.

\citet{Allison} analysed the influence of the virial ratio and clumpiness on the escape rate of stars at early times from a stellar cluster, but without including a binary population, tidal field or stellar evolution. The reason to choose not to include these features in the simulations was the short integration time of the models (roughly 4 Myr). If the clusters are to evolve for longer then all the mentioned features become important in the evolution of stellar clusters. \citet{Perets} studied the origin of early runaway stars from stellar clusters but with some simplified assumptions on the distribution of binaries. 
 We have extended these two works by e\-vol\-ving model clusters using the direct {\it{N}}-body code NBODY6 until they lose approximately 90\% of their initial $10\,000$ systems, a situation that occurs within the first 4 Gyr for all our simulations.

In our models we include the presence of the Galaxy by mo\-de\-lling it as a fixed three-component potential. The orbit that we have chosen for our clusters is circular. However, we note that the analysis of eccentric orbits, which means that clusters will travel in a changing tidal field and the mass-loss rate of the cluster changes over time \citep{Webb} is not to be discounted.
Binaries are an important component of open clusters (e.g. \citealt{delafuente97}). Recently, \cite{Geller12} derived a binary frequency of 23\% for main-sequence stars with periods less than 10$^{4}$ d in the old open cluster NGC 188. By studying nearby star forming regions, \cite{King} found that the primordial binary percentage can easily escalate to values of 70\%, assuming a period distribution similar to that found by \cite{Duquennoy} for field stars up to periods of 10$^{10}$ d. Even in the case that a star-forming region creates new stars with a primordial multiplicity of 100\%, these multiple systems (binaries, quartets, etc) will evolve to reach the primordial binary values observed in young clusters after star formation has ceased  (\citealt{Kaczmarek}; \citealt{Parker11}). Another factor to consider in the environment of a star cluster is that primordial binaries can be broken-up in encounters between stars/binaries as the cluster evolves. As a general rule there is a hard-soft binary limit where hard (close) binaries will be expected to survive long-term and soft (wide) binaries will be quickly broken-up \citep{Heggie75}. For a typical open cluster with a half-mass radius of 1 pc and N $\sim$ 10$^{4}$ the limit between hard/soft binaries is $\sim$ 40 au or a period of 10$^{5}$ d for a typical binary of 1 M$_{\odot}$. For our models we take into account different percentages of primordial binaries by focussing on values of 0\%, 10\% and 50\% in particular.

We study how stars escape from open clusters according to different criteria such as the escape velocity, stellar type at the moment 
of escape, ages of the stars escaping, the influence of primordial binaries and the mechanism responsible for these 
escapes, especially how these mechanisms may imprint different velocities on the escapers.

\section{THE MODELS}
\label{THE MODELS}

A series of {\it{N}}-body models were run of open clusters by u\-sing the code NBODY6 \citep{Aarseth3} with graphical processor unit (GPU) 
and multi-core CPU capability \citep{Nitadori}. For each of our simulations we used six CPU cores (2.66 GHz 64-bit
Intel Xeon 5650) and one Tesla C2070 GPU on the new GPU Supercomputer for Theoretical Astrophysics Research (gSTAR) at Swinburne University of
Technology.

The simulations performed consisted of stellar clusters with $10\,000$ systems\footnote{The total number of systems corresponds to the 
number of single stars plus binary systems.}. 
Single star masses between the limits of 0.1 and 50 $M_{\odot}$ were chosen 
following a Kroupa initial mass function \citep{Kroupa} 
and a metallicity typical of open clusters of Z=0.01 \citep{Friel} was used. 
We have included different primordial binary 
percentages, i.e. $0\%$, $10\%$ and $50\%$, which means that our models will have di\-ffe\-rent total masses and number of stars, as shown in Table \ref{models10k}. The masses of these binaries
were set by combining the masses of two single stars from the Kroupa initial mass function \citep{Kroupa} as the total mass of the binary system and then
 assigning the masses of the components of the system by using a uniform mass-ratio distribution. The eccentricities
 were taken from the distribution in M35 \citep{GellerIAU}. For the periods of our binaries we choose these from the distribution of 
\citet{Duquennoy}. However, we impose an upper limit of 10$^{6}$ d as very soft binaries will be quickly broken-up and not have an impact on the dynamical evolution. 
According to their initial separations, binary systems can undergo eigen-evolution, i.e. closer binaries should be
 circularized before reaching the main-sequence \citep{Kroupa2}, in which case the period distribution is modified according to \citet{Kroupa3}.
As a result we find that about 50\% of our primordial binaries are close, i.e. hard binaries 
(for each model that starts with binaries). 
These will be treated with 
the Kustaanheimo-Stiefel regularization scheme \citep{Aarseth3}.

For the positions of the stars within our cluster we have chosen a Plummer sphere model \citep{Plummer}, 
\begin{equation}
\rho(r)=\frac{3M}{4\pi r_{0}^{3}}\frac{1}{[1 + r/r_{0}^{2}]^{5/2}}. 
\end{equation}
We then obtain the velocities using the von Neumann's rejection technique, as explained in \citet{Aarseth3}. All our models are
 initially at virial equilibrium, and each of them was created by using a different random seed number to generate the positions and velocities.

Our model clusters were not evolved in isolation but within the potential of the Milky Way, which was represented by a 
 3-component Galaxy, i.e. a point mass bulge with a mass of $1.5\times10^{10} M_{\odot}$, a \citet{Miyamoto} disc with a mass of 
$5.0\times10^{10} M_{\odot}$, a softening scale length of $4\,$kpc and a vertical softening scale length of $0.8\,$kpc,
and finally 
a logarithmic halo set by requiring a circular velocity of 220 km s$^{-1}$ at a distance of $8.5\,$kpc \citep{Xue}.

The presence of the Galaxy means
that the stellar clusters will be affected by the associated tidal field so we have to decide how concentrated our clusters will be, 
keeping in mind the presence of the tidal limit which is calculated with the following relation,
\begin{equation}\label{tidal}
%r_{\rm t} = \left(\frac{M}{3M_{\rm g}}\right)^{\frac{1}{3}}R_{\rm gc},
r_{\rm t} = \left(\frac{M}{3M_{\rm g}}\right)^{1 / 3}R_{\rm gc},
\end{equation}
where the tidal radius ($r_{\rm t}$) depends on the total mass of the cluster ($M$), the mass of the
Galaxy ($M_{\rm g}$) and the distance of the cluster from the centre of its parent Galaxy ($R_{\rm gc}$).
The model clusters were placed on an orbit in the plane of the disc with an apogalactic distance of $8.5\,$kpc and a perigalactic distance of $7.2\,$kpc.
Taking $M \sim 6\,000 \, M_{\odot}$, $M_{\rm g} \sim 10^{11} \, M_{\odot}$ and $R_{\rm gc} = 8.5\,$kpc we 
get $r_{\rm t} \simeq 23\,$pc.  
We chose a scale
length for all our clusters such that they have a half-mass radius of $0.77\,$pc at the beginning of the simulations.

As a consequence of the presence of the
tidal field of the Galaxy stars will be stripped from the stellar clusters. The escape 
criterion used in all the models is to remove stars when they reach a distance greater than four times the tidal radius of the cluster 
(and have a total positive energy). In this way, we do not have to worry about potential escapers \citep{Takahashi}. Delaying the removal of stars from the simulations is preferable to premature removal before a star has fully escaped from the clutches of the cluster potential.

In this work we are interested in analysing the velocity and evolutionary stage of the escaping stars from the stellar clusters under the
 described
 tidal field, and especially the effect of the primordial binaries on these escape events. Different mechanisms are involved in the
 ejection of stars from stellar clusters
 into the Galaxy. 
The {\it{binary ejection mechanism}} ({\it{BEM}}), 
also known as the supernova ejection mechanism, 
was proposed by \citet{Blaauw} and involves stellar evolution in a 
binary system. 
The primary star of the binary system, a massive star, 
evolves more quickly than the secondary and when it reaches the supernova stage 
receives a velocity kick. 
Depending on the strength and direction of this kick the outcome can be 
either a bound binary with altered orbital parameters (which itself may escape) 
or disassociation of the binary with the secondary having an enhanced velocity 
\citep{PortegiesZwart}. 
Another mechanism is the {\it{dynamical ejection mechanism}} ({\it{DEM}}), proposed 
by \citet{Poveda}, in which dynamical interactions between two binary systems, 
or a binary system with a third star, creates in general 
one hard binary system and one (or two) runaway star(s). 
Various authors have attempted to quantify these mechanisms in terms of the magnitude 
of the velocity imprinted on the escaping star (\citealt{Blaauw}; \citealt{Poveda}; \citealt{Leonard1}; \citealt{Leonard2}; 
\citealt{Leonard3}; \citealt{PortegiesZwart}; \citealt{Gvaramadze}). 
Velocities of the order $200\,{\rm km} \, {\rm s}^{-1}$ have been shown to be typical of both 
mechanisms but values in excess of $1\,000\,{\rm km} \, {\rm s}^{-1}$ have been reported as possible for 
{\it DEM} \citep{Leonard3} and could also be obtained for {\it BEM} if suitably large velocity kicks are imparted. In practice a range of velocities is to be expected with a dependence on the masses 
involved, the position in the cluster, the strength of the interaction (for {\it DEM}) 
and the parameters of the kick velocity (for {\it BEM}). 
Finally, we do not forget the process of 
evaporation that a dynamical system undergoes, 
in which stars obtain enough energy to escape within the 
relaxation timescale owing to two-body interactions. 
This gives velocities of the order $\sqrt{2} v_\sigma$ 
\citep{BT08}, 
where $v_\sigma$ is the velocity dispersion which is typically a 
few ${\rm km} \, {\rm s}^{-1}$ for open clusters. 

Our simulations include the analytical stellar evolution approach of 
\citet[]{Hurley1,Hurley2}. 
Stellar evolution is relatively straightforward from 
the main-sequence to the red giant branch (RGB) and does not have a 
high impact on the dynamical evolution of a stellar cluster, 
although mass-loss from rapidly rotating stars on the main-sequence 
can affect the rate of escape of stars (\citealt{Maeder1}; \citealt{Meynet1} and references therein), since the tidal radius 
and potential well of the cluster depend on the total mass. Since stellar rotation is not included in the prescription for mass-loss in the current implementation of stellar evolution in NBODY6 we did not take this effect into account.
For massive stars that evolve beyond these stages life can become 
complicated. 
In particular, these stars can undergo supernova explosions 
to leave either a neutron star (NS) or black hole (BH) remnant. 
The details of the explosion, including if it is isotropic or not 
and the velocity kick that anisotropic explosion can impart, are 
far from being fully understood \citep{Lyne}.

Owing to the uncertainty in the kick treatment and the effect this 
will have on the escape velocities of stars associated with supernovae, 
we will focus mainly on escape statistics associated with {\it DEM} and evaporation. 
However, within our stellar evolution approach we still require a treatment 
of kick velocities that is as realistic as possible, particularly as 
supernovae events can alter cluster parameters and thus have a secondary 
effect on the velocities of other stars. 
In our simulations we will assume as the standard case ({\it{set1}} of models) 
a low symmetry in the supernova events that create neutron stars,  
which imprints a high velocity kick chosen at random from a Maxwellian 
distribution with a dispersion of $190\,{\rm km} \, {\rm s}^{-1}$ \citep{Hansen}
to these remnants. 
This assumption is backed by the observations
of runaway pulsars (\citealt{Blaauw0}) and hydrodynamical simulations 
(\citealt{Nordhaus} and references within). 
For the creation of black holes we will assume a lower dispersion of 
$6\, {\rm km} \, {\rm s}^{-1}$ owing to fallback of exploded material onto 
the black hole \citep{Belczynski}. We will also assume 
that the mass expelled by the stellar winds during the thermally pulsating 
asymptotic giant branch phase (TPAGB) is highly symmetric. 
Thus, the remnant of this phase of the stellar evolution, a white dwarf (WD), 
does not suffer any velocity kick. 
To attempt to quantify the effect of these assumptions on our results we will 
evolve two alternative sets of models with different choices for the velocity 
kicks, including mild kicks for white dwarfs \citep{Fregeau}. 
Details will be provided in Section \ref{KICKS}. 

We also evolve a set of models with greater initial number density and some models that have the same total mass for different binary percentages.

\begin{table*}
\begin{minipage}{110mm}
\caption{Models of 10\,000 systems arranged according to the kick assumption considered ({\it{set1}}, {\it{2}} or {\it{3}}) and the binary
 fraction used ({\it{a}}, {\it{b}} or {\it{c}}).
 The first set of models, {\it{set1a}},{\it{b}},{\it{c,}} co\-rres\-ponds to what we use as our standard model.  The second column states the number of realisations done with different random number seeds, i.e. initial conditions, 
N$_{s,0}$ is the initial number of stars, f$_{b}$ the percentage of primordial binaries, M$_{\star}$ is the total mass of each model at the beginning of the 
run in solar masses, $\rho_{c,0}$ is the initial central density in M$_{\odot}$pc$^{-3}$ (within the core radius), 
t$_{rh,0}$ the half-mass relaxation time at 
the beginning of the run and t$_{rh,2}$ the relaxation time at 2 Gyr (in Myr). 
The Kicks column shows the dispersion in km s$^{-1}$ of the Maxwellian distribution
of velocities from which the velocity kicks were chosen at random for white dwarfs (WD), neutron stars (NS) and black holes (BH). 
Details for {\it{set1}} are provided in Section~2 while the choices for 
{\it{set2}} and {\it{set3}} will be described in Section~\ref{KICKS}.
Note that $t_{rh,2}$ for each model is the average of two realisations that are identical except for the value 
of the initial seed for the random number generator (see Section~\ref{stochasticity}). 
}
\label{models10k}
\begin{tabular}{@{}lccrcrrcc}
\hline

Model & Runs & N$_{s,0}$ & f$_{b}$ & M$_{\star}$ & $\rho_{c,0}$ & t$_{rh,0}$ & t$_{rh,2}$ & Kicks(WD/NS/BH) \\

\hline

{\it{set1a}} & 2 & 10000 & 0 & 5942.3 & 4372.9 & 20.7 & 246.6 & 0/190/6\\
{\it{set1b}} & 2 & 11000 &10 & 6581.3 & 4889.8 & 21.2 & 235.7 & 0/190/6\\
{\it{set1c}} & 2 & 15000 &50 & 9121.8 & 7215.1 & 20.4 & 275.8 & 0/190/6\\

\hline

{\it{set2a}} & 2 & 10000 & 0 & 5942.3 & 4695.0 & 21.0 & 251.7 & 2/6/6 \\
{\it{set2b}} & 2 & 11000 &10 & 6581.3 & 4823.4 & 20.7 & 224.4 & 2/6/6 \\
{\it{set2c}} & 2 & 15000 &50 & 9121.8 & 7248.3 & 21.1 & 264.2 & 2/6/6 \\

\hline

{\it{set3a}} & 2 & 10000 & 0 & 5942.3 & 4652.9 & 21.0 & 229.0 & 0/190/190 \\
{\it{set3b}} & 2 & 11000 &10 & 6581.3 & 4958.7 & 21.9 & 244.8 & 0/190/190 \\
{\it{set3c}} & 2 & 15000 &50 & 9121.8 & 7074.3 & 20.7 & 262.4 & 0/190/190 \\

\hline

{\it{denser$_{a}$}} & 1 & 10000 & 0 & 6135.8 & 33456.6 & 7.2 & 202.1 & 0/190/6 \\
{\it{denser$_{b}$}} & 1 & 11000 &10 & 6547.5 & 42826.3 & 7.5 & 201.1 & 0/190/6 \\
{\it{denser$_{c}$}} & 1 & 15000 &50 & 9119.0 & 49940.7 & 7.4 & 231.5 & 0/190/6 \\

\hline
\end{tabular}
\end{minipage}
\end{table*}

\section{CLASSICAL FEATURES}

The first objective was to analyse the behaviour of the classical parameters studied 
in {\it{N}}-body simulations in the past, such as the time evolution of the 
cluster mass and the evolution of the core and half-mass radii.

\subsection{Mass-loss}

\begin{figure}
\includegraphics[angle=0,width=8.4cm]{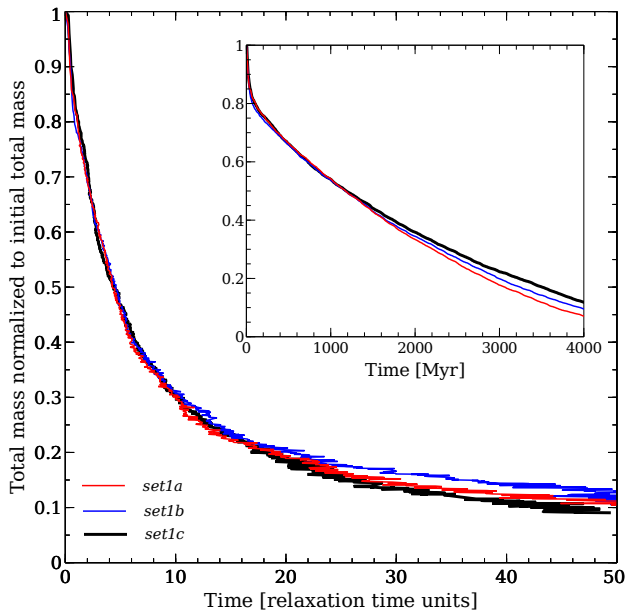}
 \caption{Mass-loss from each of our {\emph{set1}} models versus time in relaxation time units. Models with 0\% (red),
 10\% (blue) and 50\% (black) primordial binaries are shown.}
 \label{mass-loss}
\end{figure}

In Fig. \ref{mass-loss} we show the mass evolution of our {\emph{set1}} of models as a function of time scaled by the current
 half-mass re\-la\-xa\-tion time. 
As the clusters evolve they lose mass owing to tidal stripping and stellar evolution. 
It is clear from Fig. \ref{mass-loss} that the model clusters evolved by losing mass in a similar way without regards of
 the primordial binary percentage.
 That is to say that in general, the models evolved evenly on a dynamical timescale (i.e. relaxation time).
 Differences appear after 20 relaxation times but at this moment the systems have less than 20\% of their
 initial mass and fluctuations from low-number statistics are involved. We show for completeness in the same
 figure the mass-loss as a function of time in physical units (Myr). For each of the models they have lost half of their initial 
mass after 1 Gyr and subsequently differences start to appear after $\sim$ 1.5 Gyr. 
From this point on the model with 50\% primordial binaries -- which was the most massive to begin with and cor\-re\-spon\-dingly has their
 largest half-mass relaxation timescale at 2 Gyr (see Table \ref{models10k}) -- has
 more mass and as a consequence a deeper potential well, which in turn makes it more difficult for stars to escape.

\subsection{Core Radius, Core Mass and Half-Mass Radius}

The size of stellar clusters within a tidal field is determined by the interplay between the evaporation of
 stars due to the effects of the external potential field which makes the clusters shrink by removing stars and the heating produced by
 mass-loss and binaries systems in the core via two-body interactions which causes the clusters to expand 
(e.g. \citealt{Hills}; \citealt{Giersz}; \citealt{Gieles11}).
The typical information about stellar cluster sizes that both observational and computational studies take into account are the core radius 
and the half-mass radius although with some remarkable differences, e.g. observational studies used the half-light radius instead of the 
half-mass radius and the core radius is obtained by fitting a \citet{King66} profile instead of using the local stellar density \citep{Casertano}.

\begin{figure}
\includegraphics[angle=0,width=8.4cm]{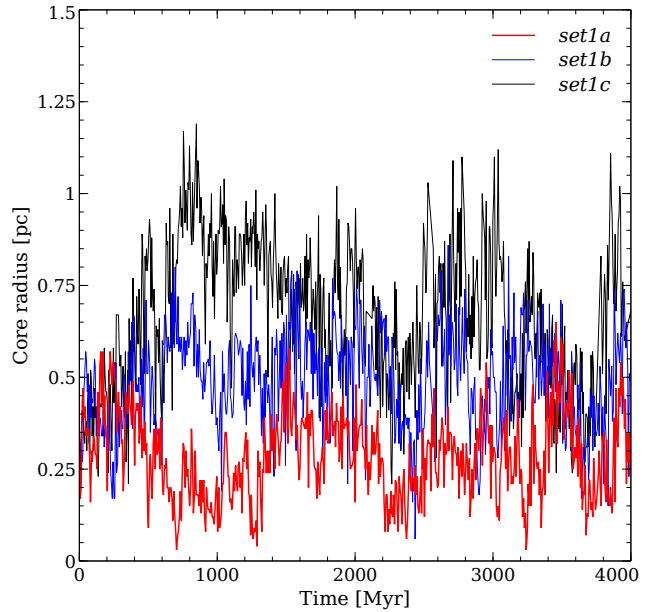}
\caption{Core radius for our standard models with different primordial binary fractions (0\% in red, 10\% in blue and 50\% in black).}
\label{core-radius1}
\end{figure}

\begin{figure}
\includegraphics[angle=0,width=8.4cm]{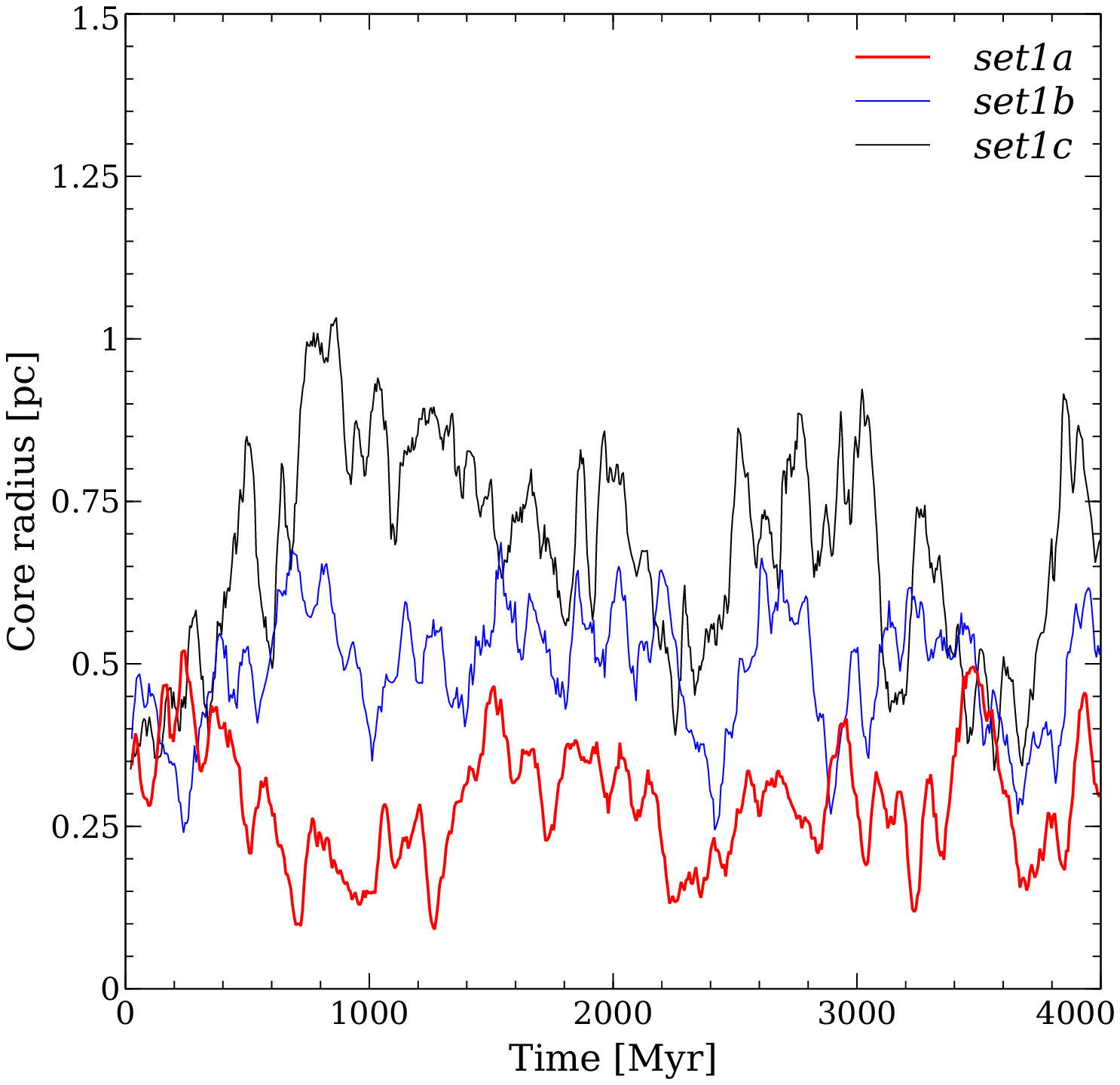}
 \caption{Core radius for our standard models with different primordial binary fractions (0\% in red, 10\% in blue and 50\% in black). In this plot we have smoothed the values of the core radius in the three curves by using a rolling average taking into account seven consecutive values.}
 \label{core-radius2}
\end{figure}

The determination of the density-weighted core radius via the method developed by \citet{Casertano} leads to a noisy 
representation that has a tendency to fluctuate markedly. 
This can make it difficult to identify the end of the core-collapse phase by naked eye inspection of the 
behaviour of the core radius with time. 
However, this method has been employed in the past, looking for the first 
global minimum that is reached by the core radius (e.g. \citealt{Fukushige1}; \citealt{Hurley04}). 
We illustrate this in Fig. \ref{core-radius1}  where we show the evolution of the core radius 
for our {\it{set1}} models. 
From this we deduce that the end of core collapse for {\it{set1a}} occurs around 0.7 Gyr 
compared to 2.2 Gyr for {\it{set1c}}. These values have been confirmed by smoothing the data (Fig. \ref{core-radius2}) using a rolling average of seven consecutive values.

Subsequent to this first discernable minimum the core then repeats the pattern of 
expansion and contraction, described as core oscillations or fluctuations 
(\citealt{Sugimoto}: see also \citealt{Heggie}). 
For {\it{set1b}} the initial core-collapse point is not so obvious but likely 
occurs at $\sim 1\,$Gyr, then repeated at $\sim 2.4\,$Gyr. 
It was expected that model {\it set1a} would reach core collapse first owing 
to the lack of an early energy source in the core (i.e. binary systems) to delay it.
In contrast, our model cluster with the highest binary percentage ({\it set1c}) undergoes an initial expansion and afterwards maintains a
 larger core radius. 
In fact, all models show some expansion initially owing to stellar evolution and then they separate in sizes, as the 
mo\-dels with primordial binaries endure a prolonged expansion phase \citep{McMillan93}.
The mass inside the core tends to be larger in the model without primordial binaries ({\it{set1a}}) since 
the segregation of heavy stars towards the centre of the cluster is more efficient \citep{Gieles10}.
We note that when core collapse is reached at 0.7 Gyr two binary systems have formed in this model and 
shed enough energy in the core to halt the collapse. 

\begin{figure}
\includegraphics[angle=0,width=8.4cm]{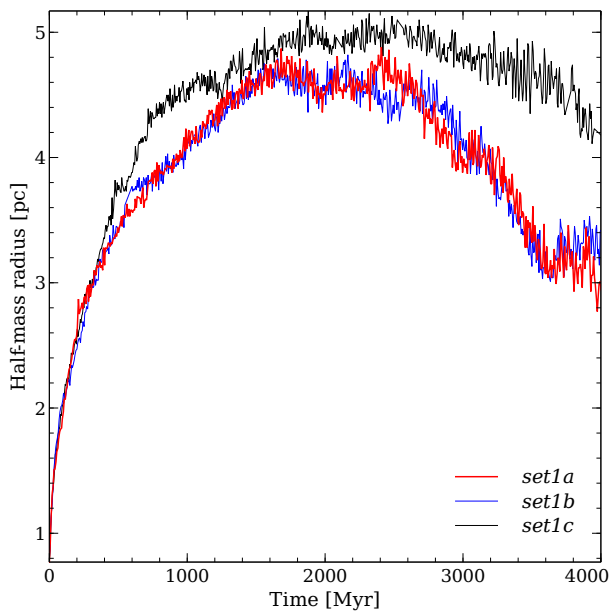}
 \caption{Half-mass radius for our standard models with different primordial binary fractions (0\% in red, 10\% in blue and 50\% in black).}
 \label{half-radius}
\end{figure}

Moving our attention to the behaviour of the half-mass radius (Fig. \ref{half-radius}), immediately noticeable is the significant expansion 
in all of the models  
during the first $\sim 1$ Gyr, fueled by mass-loss from stellar evolution initially and then maintained by two-body effects. 
The subsequent expansion has been described by \citet{Fukushige1}. 
In summary, the overall expansion of the cluster enhances the loss of stars since they will move outside the tidal
radius of the cluster. 
The effect of this loss of mass is the decrease of the tidal radius, which should lead to a decrease in the half-mass radius as well 
but the density of stars near the tidal radius is not high as a consequence of mass segregation.
So what is really seen in practice is a further expansion
 of the cluster, i.e. a further increase in the half-mass radius. 
Eventually, tidal interaction wins the battle at $\sim 2$ Gyr for our models 
and we see a turnover in the half-mass radius, as has been shown previously for models with a tidal field 
(e.g. \citealt{Giersz}). 
The half-mass radius then starts to shrink owing to an enhancement of escaping stars. 
For the model with $50\%$ primordial 
binaries ({\it{set1c}}) this turnover in the half-mass radius is less pronounced.

\subsection{Velocity Dispersion}

\begin{figure}
\includegraphics[angle=0,width=8.4cm]{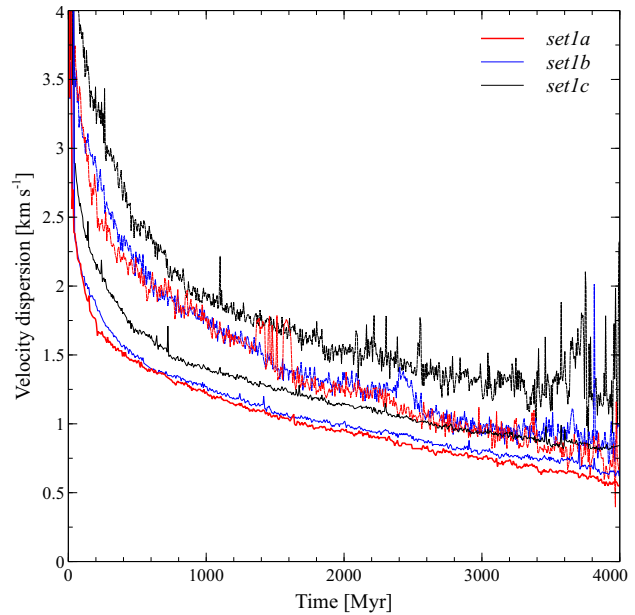}
 \caption{Velocity dispersion for our standard models with different primordial binary fractions (0\% in red, 
10\% in blue and 50\% in black). The three upper dotted curves represent the velocity dispersion within the 10\% Lagangrian radii.}
 \label{vdispersion}
\end{figure}

Fig. \ref{vdispersion} shows the time evolution of the velocity dispersion in our {\it{set1}} of models. The most distinguished feature is the 
drop that all models show at the beginning of the evolution which is a consequence of the overall expansion that each model undergoes 
as a result of stellar evolution mass-loss. 
Mass segregation also plays a role, causing the most massive stars 
to move towards the centre of the cluster and owing to equipartition of energy to transfer their kinetic energy to the low mass stars \citep{Chernoff}.
 These low mass stars in turn will move outside the limit of the cluster and the dynamics of the core will be dominated by more massive stars with low
 rms velocities. All the curves in the figure show ``spikes'', which are caused by high velocity escapers \citep{Giersz}. The larger values for 
the velocity dispersion in {\it{set1c}} are a direct result of the higher number of binary systems in this model 
which heat the central regions through three- and four-body encounters \citep{Gieles10b}.
In Fig. \ref{vdispersion} we also show the velocity dispersion within the 10\% Lagrangian radii. 
As expected, the velocity dispersion is in general higher in the inner regions, representing the higher stellar 
densities which lead to a greater incidence of dynamical encounters. 
Once again the presence of a substantial population of binaries increases the velocity dispersion 
(or temperature: see \citealt{Giersz} and \citealt{Kupper} for example). 

\subsection{Stochasticity/sampling validation}
\label{stochasticity}

In order to validate our results we have run simulations with different seed numbers, i.e. different realisations of the initial conditions, and checked the impact of stochasticity in our models. 

We have confirmed the times where core collapse occurs for our {\it{set1}} of models. The sequence of increasing core collapse times is maintained across the models with increasing binary percentages, although the behaviour of {\it{set1b}} is still erratic, showing similarities with {\it{set1a}}. Both {\it{set1a}} and  {\it{set1b}} are clearly distinguished from {\it{set1c}}. The half-mass radius still ranges between 3 to 5 pc after 2 Gyr (Fig. \ref{half-radius}) depending on the primordial binary fraction of the cluster. The only difference was an amplified increase of the half-mass radius for the new realisation of the model with 50\% of binaries at around 1 Gyr owing to the formation of additional hard binaries in the core. By $\sim3.5$ Gyr the difference has disappeared and the two realisations continued with a similar half-mass radius. The velocity dispersion is also similar between the realisations, only showing different spikes for the models with 50\% of binaries which represent that high velocity escaper events occur at different times.

Where applicable we have utilised the second realisations to strengthen our analysis. Such instances will be clearly specified.

\section{Fugitives from Stellar Clusters}

We have discussed the structure of our model clusters and the global effect that the loss of mass by both stellar evolution and star escapes imprints on 
their evolution and dominant lifetime, but the question remains: which are the mechanisms that force stars to abandon their parent cluster?

As mentioned in Section \ref{THE MODELS}, two main mechanisms have been proposed to explain the ejection of members from a stellar cluster: {\it{the binary ejection me\-cha\-nism}}
 ({\it{BEM}})
and {\it{the dynamical ejection me\-cha\-nism}} ({\it{DEM}}).

There has been fruitful discussion in the literature about whether one of these mechanism is dominant (\citealt[]{Leonard1,Leonard2}) or
 whether some different mechanism is working to produce stars that gain velocities higher than the escape velocity of their parent cluster, e.g. collisions
 with high mass stars (masses higher than 50 $M_{\odot}$: \citealt{Gvaramadze}).

It is clear now\-a\-days that neither of the mentioned me\-cha\-nisms alone will be responsible for producing escapers. Instead it will be some combination of them.
In this work we have focused our attention on the impact that the primordial binary fraction will have on the escape
 rate of stars from stellar clusters and on the overall distribution in phase space coordinates of the escapers, as well as a breakdown of statistics by stellar evolutionary stage. 

\subsection{Impact of the primordial binary fraction on the escape velocity}

\subsubsection{Single stars}
\label{single}

\begin{figure}
\includegraphics[angle=0,width=8.4cm]{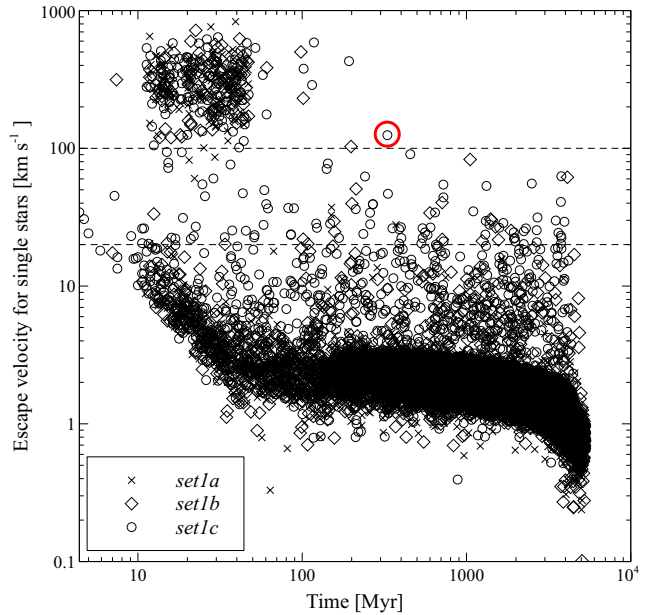}
\caption{Time evolution of the escape velocity of single stars. The horizontal lines represent the limits between the three escaper groups: normal-, high- and hyper-velocities. The red circled dot corresponds to a 3 M$_{\odot}$ star that originally would have evolved into a white dwarf, but during its evolution it formed a binary system that due to three-body encounters hardened enough to merge, providing enough mass to the original 3 M$_{\odot}$ star to evolve into a neutron star and eventually escape with hyper-velocity. %There are several neutron stars with velocities between $200$-$800$ km s$^{-1}$ that are not diplayed for the sake of the scale, but we will discuss them in Section \ref{DISCUSSION}.}
There rest of the escapers with velocities between $200$-$800$ km s$^{-1}$ are neutron stars. We will discuss them in Section \ref{DISCUSSION}.}
\label{velsintime}
\end{figure}

In Fig. \ref{velsintime} data from the {\it{set1}} models of various primordial binary fraction are combined to show the escape velocity of single stars across the 4 Gyr of evolution of the clusters studied. The criterium used to identify an escaper was to reach a distance of 100 pc (approximately four times the tidal radius of our models) and, of course, to have a positive total energy, i.e. unbound.
It is easy in theory to differentiate the three main
 process that are responsible for driving escapers, although it is harder to define limits between them. For our purposes we distinguish the following three regimes:

\begin{itemize}
\item[$\blacktriangleright$] Hyper-velocity escapers ({\it{v $\geq$ 100 km s$^{-1}$}}), more likely to come from early stellar evolution process (i.e. supernovae), but few body interactions could provide hyper-velocity escapers as well (\citealt{Napiwotzki}),
\item[$\blacktriangleright$] High-velocity escapers ({\it{20 km s$^{-1}$ $\leq$ v $\leq$ 100 km s$^{-1}$}}), more likely to come from few body interactions and binary disruption (\citealt{Poveda}; \citealt{Royer}; \citealt{Berger}; \citealt{Hoogerwerf}),
\item[$\blacktriangleright$] Normal-velocity escapers ({\it{v $\leq$ 20 km s$^{-1}$}}), more likely to come from evaporation.
\end{itemize}

\begin{figure}
\includegraphics[angle=0,width=8.4cm]{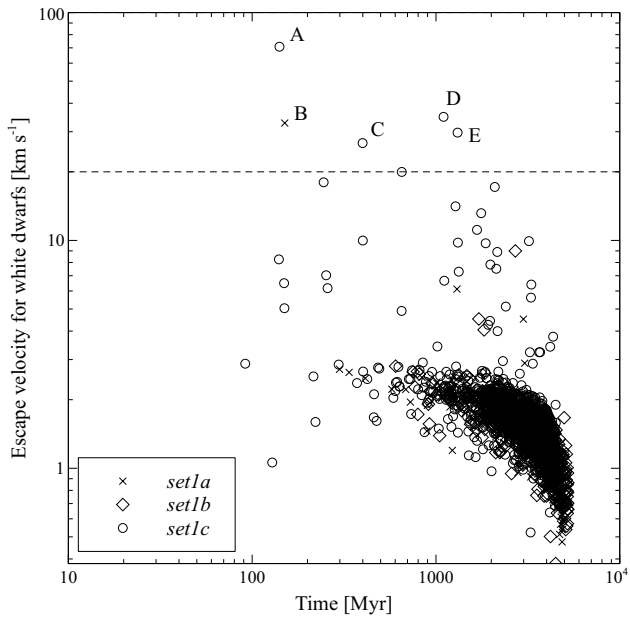}
 \caption{Time evolution of the escape velocity for white dwarfs. The horizontal line represent the limit between the normal and high velocity escaper groups.}
 \label{velsintimeWD}
\end{figure}

In all the cases within our standard models $99\%$ of the escapers correspond to the normal-velocity group, which means that the bulk of our escapers will be a byproduct of evaporation and low energy two-body encounters. All the hyper-velocity escapers in the standard models are neutron stars.

We proceed with further analysis of our high-velocity escapers. There exists a large quantity of information on high-velocity stars within our Galaxy (\citealt{Berger}; \citealt{Hoogerwerf}; \citealt{Li}; \citealt{Royer}; \citealt{Napiwotzki}; \citealt{Nidever} and references therein), and even more speculation about the origin of these stars (\citealt{Blaauw}; \citealt{Poveda}; \citealt{PortegiesZwart}; \citealt{Gvaramadze}; \citealt{Oh}). One of the main problems with identifying the birth places of runaway stars is the necessity to obtain accurate spectroscopic information, which is essential for calculating the evolutionary stage of the runaway star. If the latter is known (or can at least be assumed with some confidence) then the kinematics of the stars can be used to constrain the distance travelled by these escapers and henceforth their birth location, which potentially can be a star cluster \citep{delaFuente}. Spectroscopic information is not always easy to obtain for halo stars, so having statistics on the type of stars that can reach the halo from model stellar clusters within the disc can help with these constraints.

With this in mind we can utilise our simulations to put initial constraints on the distances that runaway stars from star clusters can travel. In particular we start by focussing on the main-sequence stars that escape from clusters and ask if these stars can reach the halo of the Galaxy as main-sequence stars or giants.

Table \ref{escap} includes all high-velocity stars that have escaped as main-sequence stars with masses greater than 1 M$_{\odot}$, as well as a few giants that escape (from all the realisations of our {\it{set1}} models). The information in Table \ref{escap} describes the range of times in which we could observe these stars as main-sequence stars and/or runaway giants in the Galactic field. Taking an inner limit of the distance to the dark matter halo to be 20 kpc, we find that a range of stars (with masses < 2.4 M$_{\odot}$) reach the halo as either a main-sequence star or a giant, potentially to be observed (e.g. \citealt{Schuster}). Note that the distances that we show in Table \ref{escap} only take into account the velocity of the stars with respect to the centre of mass of the modelled stellar clusters and not the orbital velocity of the clusters. In section \ref{DISCUSSION} we will discuss the implications of adding the orbital information around the centre of mass of the Galaxy.

\begin{table*}
\caption{Timescales and distances for main-sequence stars that escape from all the realisations of our {\it{set1}} models 
with mass greater than 1 M$_{\odot}$ and high escape velocity, sorted according to the mass of the stars at the moment of escape. The escape velocity is in km s$^{-1}$. $T_{esc}$ is the time when the star leaves the cluster, $T_{evol}$ is its main-sequence lifetime and $T_{wd}$ indicates the time when the star will become a white dwarf (and will no longer be visible as a giant). All times in Myr. Next we have distances in kpc travelled by these stars after leaving the cluster. The column DistMS is the total distance that escapers will travel as a main-sequence star before evolving into a giant and DistGiant is the total distance travelled after becoming a giant and before evolving into a white dwarf. We are assuming that these stars will travel in a straight line without following the potential of the Galaxy. $^{\ddagger}$These particular stars escaped as giants.
$^{\diamond}$These stars were originally in binary systems and subsequently merged with their corresponding companion, mixing in new material and altering their evolution timescales. The mass inside the parenthesis indicates the original mass of these stars before the merger. 
For the main-sequence lifetime we have combined the time that the un-mixed star lived on the main-sequence before the merger with the remaining main-sequence lifetime of the newly formed star. The giant lifetime is based solely on the mass of the merged star.
$^{\ast}$During its lifetime this star was involved in several exchanges in binary systems to finally engulf a white dwarf companion just prior to escape at 598 Myr. We could infer that this new star will only live a further 1 Myr as a giant before evolving to finally become a white dwarf. The actual distance travelled by this star as a giant is 0.03 kpc.
$^{\bullet}$These entries are stars that will become a neutron star and not a white dwarf, i.e. $T_{wd}$ corresponds to the time when the star becomes a neutron star.
For timescales of stars after they have left the cluster, eg. $T_{evol}$ and $T_{wd}$, we use the Single Star Evolution (SSE) package of \citet{Hurley1} which is the same stellar evolution prescription as used internally in NBODY6.}
\label{escap}
\begin{tabular}{@{}lrrrrrr}
\hline
Mass & Velocity & $T_{esc}$ & $T_{evol}$ & $T_{wd}$ & DistMS & DistGiant\\
\hline

1.003 & 25.0 & 2235.2 & 8982.7 & 10193.0 & 164.9 & 29.6\\

1.056 & 40.8 & 3754.3 & 7431.4 & 8509.1 & 146.7 & 42.9\\

1.059 & 23.6 & 579.3 & 7355.3 & 8426.2 & 156.4 & 24.7\\

1.080 & 49.7 & 58.8 & 6852.0 & 7877.7 & 330.1 & 49.8\\

1.127 & 26.3 & 354.1 & 5890.1 & 6823.9 & 142.4 & 24.0\\

1.132 & 47.2 & 43.5 & 5799.4 & 6719.8 & 265.7 & 42.5\\

1.277 & 28.1 & 259.3 & 3851.6 & 4474.6 & 98.7 & 17.1\\

1.346 & 20.4 & 531.2 & 3268.9 & 3772.8 & 54.6 & 10.1\\

1.373 & 25.3 & 1867.0 & 3080.5 & 3541.3 & 30.0 & 11.4\\

1.381 & 90.6 & 454.8 & 3027.5 & 3476.7 & 227.9 & 39.8\\

1.420 & 25.4 & 2675.3 & 2786.6 & 3185.3 & 2.8 & 9.9\\

1.455$^{\ddagger}$ & 31.4 & 2704.3 & 2592.5 & 2953.3 & --- & 11.1\\

1.487 & 24.5 & 1943.6 & 2431.1 & 2762.6 & 11.7 & 7.9\\

1.514 & 36.6 & 61.9 & 2305.7 & 2615.6 & 80.3 & 11.1\\

1.584 & 30.6 & 4.7 & 2019.9 & 2285.2 & 60.3 & 7.9\\

1.595 (1.227)$^{\diamond}$ & 26.8 & 1503.3 & 1961.6 & 2218.6 & 12.0 & 6.7\\

1.617 & 22.4 & 167.4 & 1902.3 & 2151.1 & 37.9 & 5.4\\

1.663$^{\ddagger}$ & 21.1 & 1984.0 & 1753.9 & 1984.0 & --- & 4.7\\

1.700 & 21.9 & 101.5 & 1646.1 & 1864.7 & 33.1 & 4.7\\

1.806 (0.974)$^{\diamond}$ & 23.2 & 3755.2 & 4619.3 & 6225.4 & 19.6 & 4.7\\

1.943 & 44.0 & 723.5 & 1124.7 & 1453.6 & 17.3 & 14.2\\

1.966 & 21.8 & 712.2 & 1087.9 & 1407.1 & 8.0 & 6.8\\

1.975 & 54.5 & 723.0 & 1074.0 & 1389.5 & 18.7 & 16.8\\

2.185 & 22.3 & 401.4 & 810.4 & 1051.0 & 8.9 & 5.2\\

2.188 & 26.6 & 603.4 & 807.4 & 1047.0 & 5.3 & 6.2\\

2.198 & 27.8 & 58.2 & 797.3 & 1033.8 & 20.1 & 6.4\\

2.185 (1.313)$^{\diamond}$ & 20.8 & 2675.9 & 2859.1 & 3890.3 & 3.7 & 4.8\\

2.249 & 36.9 & 82.3 & 748.7 & 970.1 & 24.0 & 7.9\\

2.397 & 37.3 & 150.0 & 629.7 & 812.5 & 17.5 & 6.7\\

2.674$^{\ddagger}$ & 24.5 & 559.3 & 470.4 & 599.4 & --- & 3.1\\

2.766 (2.153)$^{\ast}$ & 33.9 & 598.0 & --- & 599.0 & --- & 0.0\\

2.926 & 29.5 & 156.2 & 371.8 & 466.7 & 6.2 & 2.7\\

2.954 & 23.8 & 85.4 & 362.8 & 454.5 & 6.5 & 2.1\\

3.721 & 28.4 & 147.9 & 203.1 & 244.3 & 1.5 & 1.1\\

3.842 & 25.3 & 37.5 & 187.9 & 224.8 & 3.7 & 0.9\\

3.871 & 44.8 & 25.0 & 184.5 & 220.5 & 6.9 & 1.6\\

3.909 & 54.8 & 24.7 & 180.2 & 215.4 & 8.3 & 1.9\\

5.488 (4.778)$^{\diamond}$ & 39.3 & 60.0 & 81.8 & 176.4 & 0.8 & 0.5\\

5.827 & 22.5 & 16.0 & 72.5 & 83.2 & 1.2 & 0.2\\

5.984 & 22.5 & 51.0 & 68.5 & 78.5 & 0.4 & 0.2\\

6.210 (4.157)$^{\diamond}$ & 25.2 & 45.4 & 81.9 & 154.6 & 0.1 & 0.2\\

6.322 & 20.5 & 15.1 & 61.0 & 69.8 & 0.9 & 0.2\\

6.928 & 20.0 & 33.9 & 50.6 & 57.7 & 0.3 & 0.1\\

7.957$^{\bullet}$ & 27.0 & 22.5 & 38.6 & (43.4) & 0.3 & 0.1\\

8.367$^{\bullet}$ & 23.1 & 21.9 & 35.2 & (39.4) & 0.3 & 0.1\\

\hline
\end{tabular}
\end{table*}

Fig. \ref{velsintimeWD} provides a focus on the escape velocity of white dwarfs in our models. We follow the history of each of the high velocity white dwarfs that escape:

\begin{itemize}
\item[A -]This 1 M$_{\odot}$ white dwarf began its life as a 5 M$_{\odot}$ star in a primordial binary system with a smaller companion (3.5 M$_{\odot}$). First it merged with its companion and then when it evolved to a white dwarf it formed a new binary system with a black hole of 11 M$_{\odot}$. Finally this binary system suffered a three-body encounter with a black hole of $\sim$ 10 M$_{\odot}$ which resulted in the formation of a BH-BH binary and the ejection of the white dwarf.

\item[B -]This high-velocity escaper (initally a star of 3.7 M$_{\odot}$, but now a white dwarf of 0.9 M$_{\odot}$) was in  a binary system that involved a black hole. This binary system interacted with a single star in a three-body encounter, with the outcome being the ejection of the white dwarf.
\item[C -]This 1.0 M$_{\odot}$ white dwarf began as a single star of 4.9 M$_{\odot}$ and was ejected in a two body encounter with a 6 M$_{\odot}$ black hole.
\item[D -]This 0.7 M$_{\odot}$ white dwarf began as a 2.1 M$_{\odot}$ single star and soon after combined with a 1.3 M$_{\odot}$ star. This binary system interacts with another binary system with the outcome being the ejection of the star in question.
\item[E -]This 0.7 M$_{\odot}$ star began as a 2.1 M$_{\odot}$ single star which formed a binary system early in its lifetime. It has a promiscuous life being part of different binary systems, i.e. it is involved in the exchange of companions in three-body encounters, until it is finally ejected when its companion (a black hole of 9.8 M$_{\odot}$) formed a new binary with a giant star of $\sim$ 2 M$_{\odot}$.
\end{itemize}

The occurrence of these high-velocity WDs also raises the possibility of WDs escaping from open clusters making a contribution to the population of WDs in the Galactic halo (e.g. \citealt{Spagna}). Noting that in our standard models we have not included a velocity kick at the time of birth for white dwarfs, we have to keep in mind that the velocity distribution of the white dwarfs that escape from our simulations can be influenced by the assumptions at the moment of creation,
i.e. low/high symmetry in the ejection of stellar winds during the late phases of the asymptotic giant branch (AGB). In section \ref{KICKS} we explore the impact of these assumptions.

\subsubsection{Binary stars}

The energy budget within a stellar cluster is dominated by the energy locked in binary systems. Binaries can delay or even halt core collapse \citep{Hut}. For clusters immersed in the tidal field of their parent Galaxy, binaries can accelerate the disruption rate by injecting energy mainly in the core of the stellar system which causes the cluster to expand within the tidal radius, a process that favours stars in the outskirts of the cluster to escape through evaporation.

We analysed the impact of binaries on our escape rate of stars from the modelled clusters by selecting three possible scenarios, i.e. primordial binary percentages of 0\%, 10\% and 50\%.

\begin{figure}
\includegraphics[angle=0,width=8.4cm]{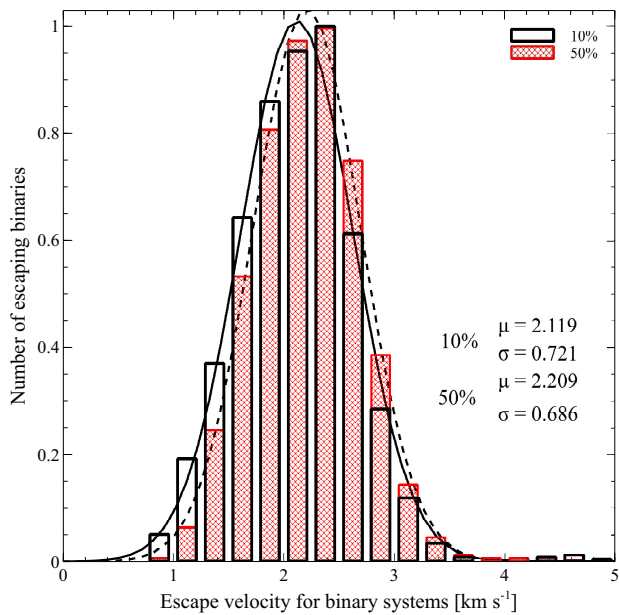}
 \caption{Velocity distribution of binary escapers for the models with 10\% and 50\% of primordial binaries. The histograms for each plot are the average of two realisations. The solid line corresponds to the gaussian fit for {\it{set1b}} and the dotted one for {\it{set1c}}. The median and dispersion of each fit are shown as well. The high-velocity tail is not displayed for the sake of the scale.}
 \label{velbinhist}
\end{figure}

In Fig. \ref{velbinhist} we show the normalised distribution of escaping binary systems as a function of their escape velocity for both {\it{set1b}} and {\it{set1c}}. We also fit a Gaussian to each histogram.

We notice that the distribution of {\it{set1c}} is skewed slightly towards higher velocities. This was expected since the higher number of binaries in this model will favour a larger number of two-body interactions (binary-single or binary-binary) which in turn causes the binaries to increase their kinematic energy as they harden in response to surviving the interaction (or disrupt if they are weakly bound) and increase the velocity dispersion (see Fig. \ref{vdispersion}). The Gaussian fits quantify the difference, showing that the mean cdshifts from $\mu = 2.1$ to $\mu = 2.2\, {\rm km}\, {\rm s}^{-1}$ as the primordial binary percentage increases and the dispersion of the Gaussian decreases.

\begin{figure}
\includegraphics[angle=0,width=8.4cm]{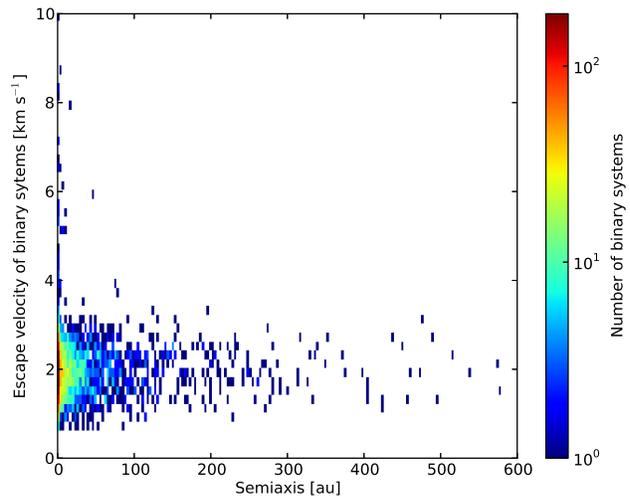}
 \caption{Escape velocity vs semiaxis for the models with 10\% and 50\% of primordial binaries. The model without primordial binaries, i.e. {\it{set1a}}, only contained
about 10 binary systems over its lifetime and is not represented
in the figure.}
 \label{velbindots}
\end{figure}

In Fig. \ref{velbindots} we show the escape velocity as a function of the semi-major axes of the binary systems that are escaping. In the figure we only show the escaping binaries with velocities below $10\, \rm{km}\, \rm{s}^{-1}$ and semi-major axes below 600 au. We note that the complete sample of binaries
 that are escaping includes four hard binary systems with velocities near $20\, \rm{km}\, \rm{s}^{-1}$ and one hard binary escaping with a velocity equal to $100\, \rm{km}\, \rm{s}^{-1}$. Also some broader binaries (21 systems), with semi-major axes ranging between 600 to 1650 au. These are not shown for the sake of the scale. 

Compared to single stars (as shown in Fig. \ref{velsintime} and discussed in Section \ref{single}) we find no hyper-velocity binary escapers and the incidence of high-velocity escapers is much less. This can be explained by keeping in mind that binary systems are more massive on average than single stars. Since the binary systems are ``heavier'' the ejection process will imprint lower velocities on them. We can see that the bulk of the escaping binaries will have short separations, which implies that most of these binaries are hard at the moment of escape. Soft binaries will escape the clusters as well but are more likely to be disrupted before leaving the cluster. We do not see any clear correlation between the escape velocity and the orbital properties of the binaries.

\subsection{Nature of escapers}
\label{nature}

\begin{figure*}
\includegraphics[angle=0,width=15.0cm]{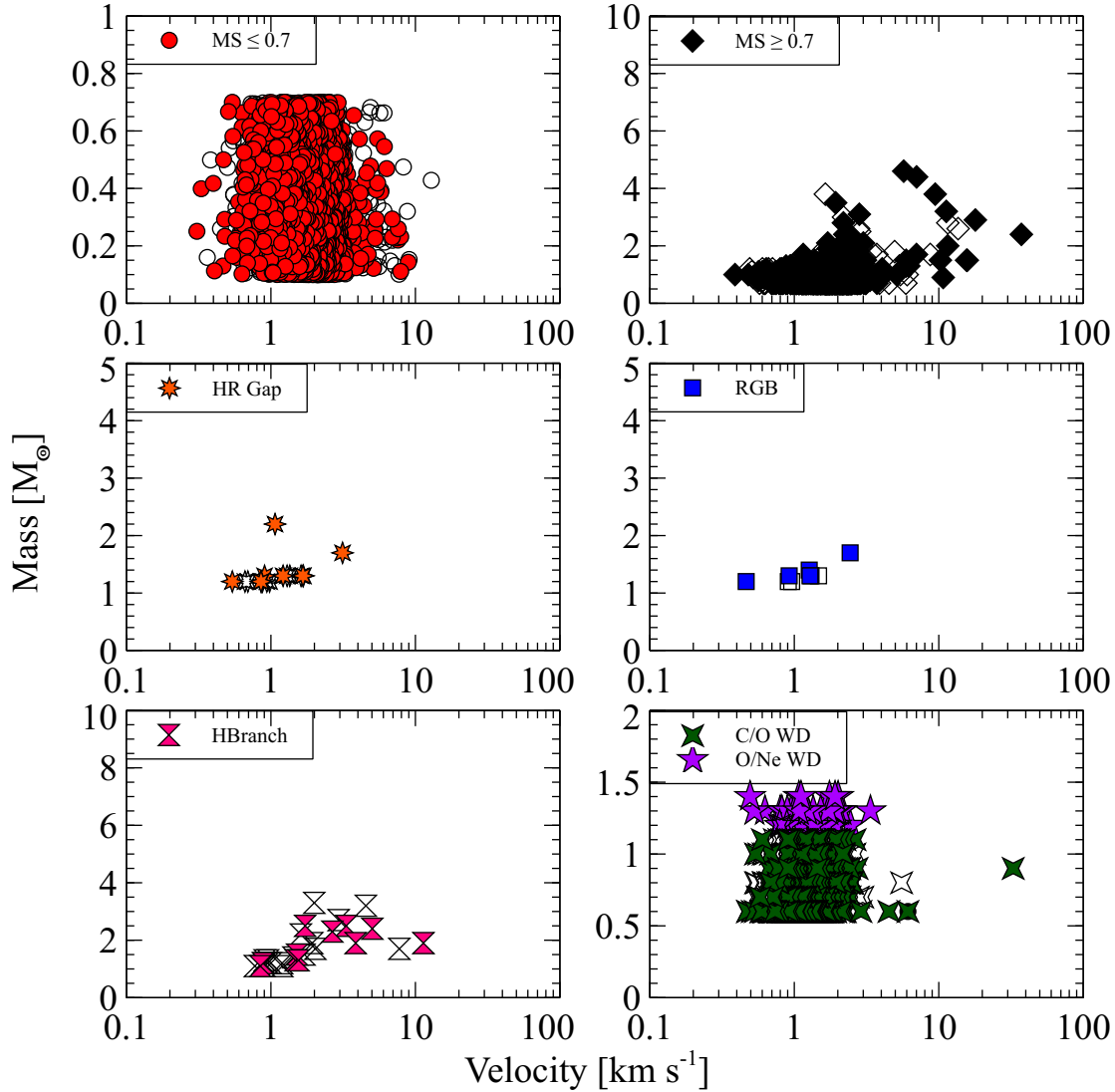}
\caption{Mass escaping as a function of the escape velocity for our {\it{set1a}}. The upper panels represent main-sequence (MS) stars for our two mass ranges: low mass (M<0.7M$_{\odot}$) and high mass (M>0.7M$_{\odot}$). The middle panels represent stars becoming sub-giant (left panel) and red giant stars (right panel), noting that in these models no asymptotic giant branch (AGB) stars are escaping but some of them are present in the clusters at different times as can be seen in Table \ref{census}. The left bottom panel are stars burning He in their core at the moment of escape. Finally, the right bottom panel represent the population of Carbon-Oxygen (C/O) and Oxygen-Neon (O/Ne) white dwarfs escaping. The filled and open symbols represent the two realisations made for each model.}
\label{split00}
\end{figure*}

\begin{figure*}
\includegraphics[angle=0,width=15.0cm]{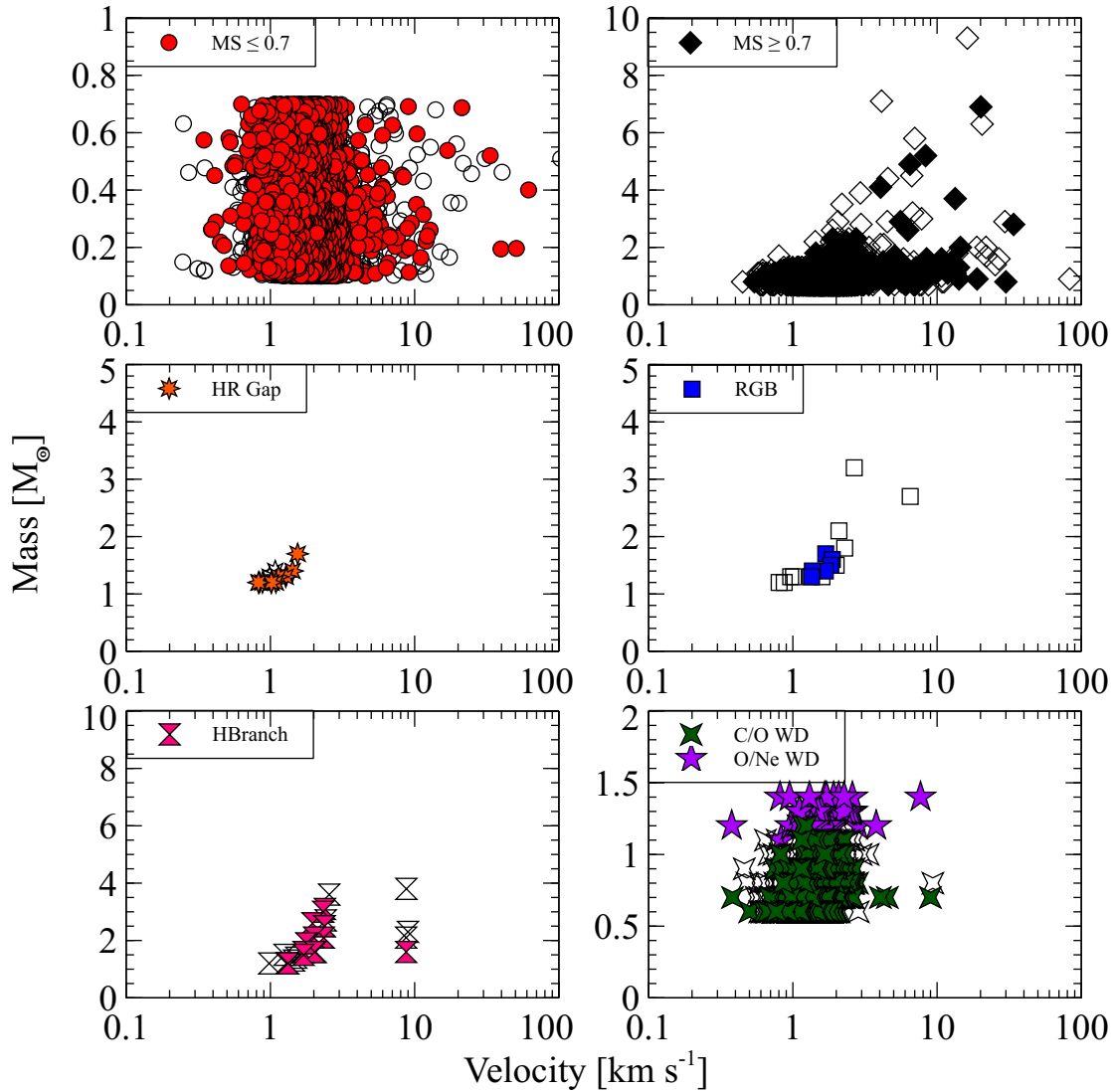}
\caption{As Fig. \ref{split00} but for {\it{set1b}}.}
\label{split10}
\end{figure*}

\begin{figure*}
\includegraphics[angle=0,width=15.0cm]{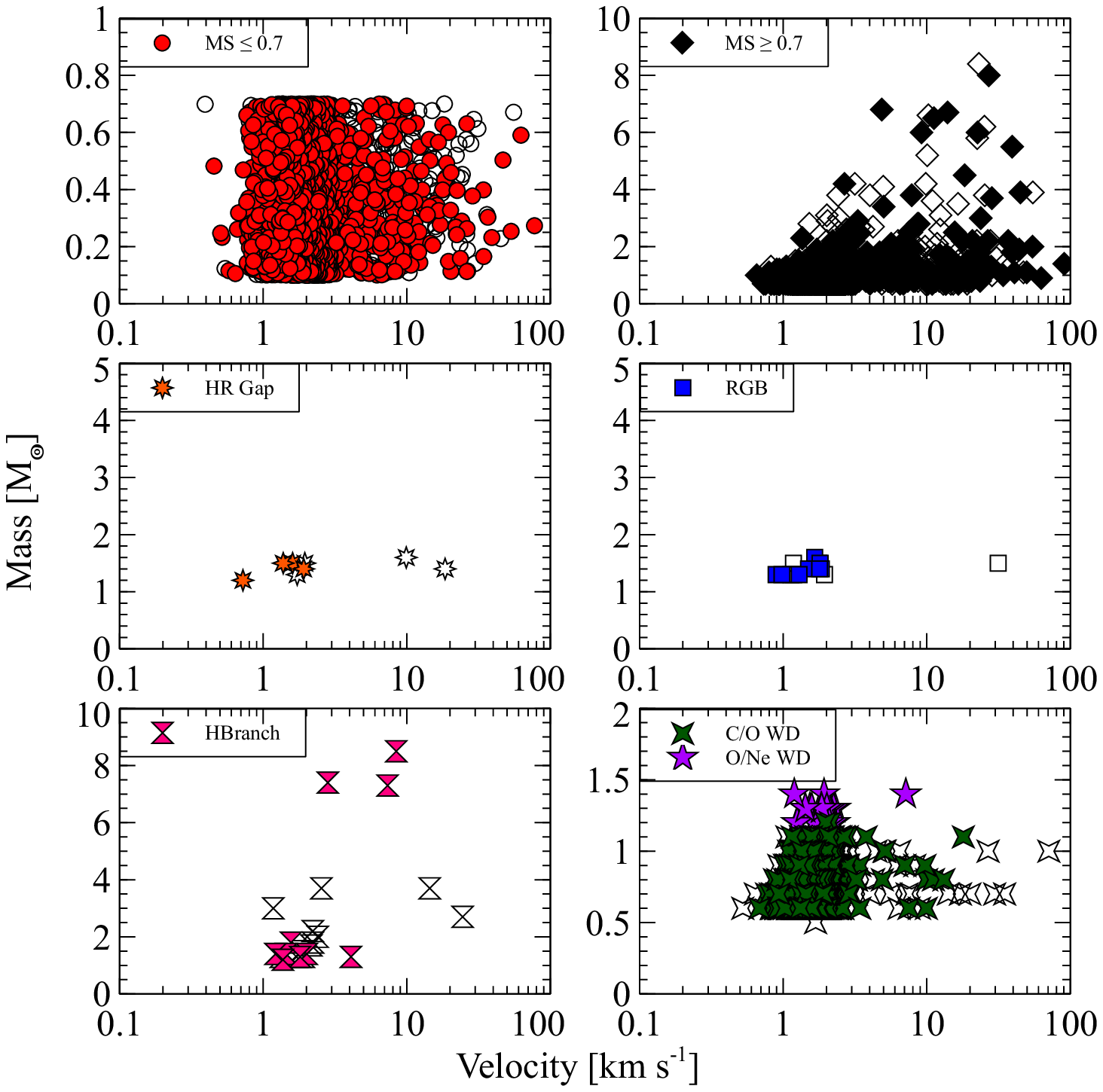}
\caption{As Fig. \ref{split00} but for {\it{set1c}}.}
\label{split50}
\end{figure*}

All the escapers that we have analysed in the previous sections will become a part of the population of the Galaxy, which leads to the general question: what type of stars will
 escape the stellar clusters to populate the galactic disc and halo? The answer to this question will be affected by our assumptions on the
 initial mass function (IMF), the stellar evolution model and velocity kick (if any) to white dwarfs, neutron stars and black holes used at the moment of
 creation of these remnants. Starting with our standard models, in Figs. \ref{split00} to \ref{split50}  we quantify the stars escaping from these models according to their evolutionary stage.

As we can see from these figures and also Table \ref{census}, the two main groups escaping are main-sequence stars, preferentially those with masses below 0.7 M$_{\odot}$, and C/O white dwarfs.
 The former is a byproduct of the interplay between mass segregation and early stellar evolution, which are processes that redistribute energy that preferentially
 will be gained by low mass stars due to energy equipartition (\citealt{Aarseth3} and references therein). The fact that low-mass stars are the most abundant from the IMF also plays a role in the census of escapers.

It is evident in Figs. \ref{split00}\,-\,\ref{split50} that from our simulations the majority of the high-velocity escapers are main-sequence stars, as previously highlighted in sec \ref{single} and Table \ref{escap}.
Of main interest are the stars with Sun-like
masses or higher, since these can became what observers have defined as runaway OB stars.

Some sub-giant and giant stars on the Hertzsprung-Russell (HR) gap or first giant branch do leave the model clusters through evaporation (all with velocities < $4\, \rm{km}\, \rm{s}^{-1}$) but we do not get any high-velocity escapers of these types.

Stars on the horizontal branch (or blue loop) also contribute to the normal velocity group through evaporation although these include a number with velocities in the range $5-15\, \rm{km}\, \rm{s}^{-1}$ produced by close encounters. There is also one high-velocity escaper ($24.5\, \rm{km}\, \rm{s}^{-1}$) of horizontal branch type from {\it{set1c}}.
This particular star belonged to a binary system that suffered a binary-binary interaction resulting in the high-velocity ejection. We do not find any asymptotic giant branch escapers. With the sub-giants and giants, the relative evolution timescales are a factor. The asymptotic giant branch phase is brief which makes it statistically unlikely to find an escaper of this type, whereas the horizontal branch phase is longer than the HR gap or RGB phases which provides more time for a close encounter to occur.

Previously in Section \ref{single} and Fig. \ref{velsintimeWD} we highlighted the origin of the high-velocity white dwarfs escaping from our models.
 These high-velocity white dwarfs support the dynamical ejection mechanism \citep{Poveda}. It is important to note that there is a delay of $\sim 1$ Gyr in the onset of the escape of white dwarfs in comparison with the rest of the stars. This is expected since in order for a white dwarf to escape, first it has to be created.
The fact that WDs are more likely to be created in the centre of the cluster, evolving from the most massive main-sequence stars at the time enhances the possibility of a high-velocity ejection.

We have performed a census of the population of stars within the cluster and those that escape in Table \ref{census}. As we mentioned in Section \ref{THE MODELS} approximately 50\% of the primordial binaries of each model will be hard binaries, which means that owing to two-body encounters half of the original primordial binary population will be more likely disrupted and the other half will likely increase their binding energy, tightening even more \citep{Hut}. The sudden drop of the number of main-sequence stars in binary systems in the first Gyr for the models with primordial binaries is a clear representation of this early two-body process. 

The higher number of white dwarfs stars in the model with 50\% primordial binaries with respect to the number in our model without primordial binaries at 4 Gyr (55\% more white dwarfs) is representative of the way in which binary systems are constructed. Binary systems are created by combining the mass of two stars taken from the Kroupa-IMF \citep{Kroupa} and then redistributing the masses of the components of the system by using a uniform mass-ratio distribution. This redistribution of mass will favour high-mass main-sequence stars \citep{Kroupa} that will likely evolve into white dwarfs.

Following the census of escapers for binary systems, it can be seen
from Table 4 that $\sim$ 60\% of the initial number of binary systems will
escape from each model that contained primordial binaries, i.e. {\it{set1b}}
and {\it{set1c}}. The overall percentage of escapers (single stars plus binaries) that
are binary systems is 6\% for {\it{set1b}} (compared to an initial cluster
binary content of 10\%) and 23\% for {\it{set1c}} (compared to 50\%) across the
4 Gyr of cluster evolution studied.
The binary fraction within the clusters at 4 Gyr is 8\% and 27\% for
{\it{set1b}} and {\it{set1c}}, respectively, with the remaining binaries having
been broken-up in encounters or the subject of mergers.

\begin{figure}
\includegraphics[angle=0,width=8.4cm]{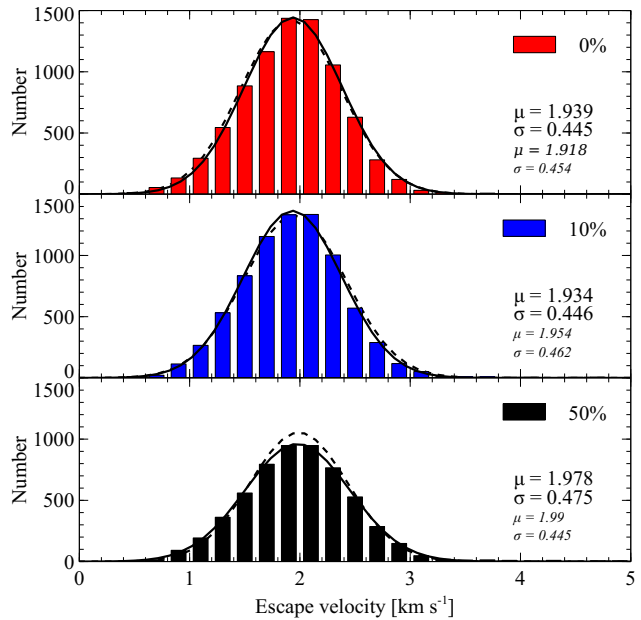}
\caption{Histograms of the escape velocity for stars in our standard models with masses below 0.7 M$_{\odot}$ for the three primordial binaries models (top panel: 0\%, mid panel: 10\% and bottom panel: 50\%). Each histogram is the average of two realisations. The solid line represents the Gaussian fit done for the first realisation with the corresponding median and dispersion at the right of each plot. The dotted line is the Gaussian fit done to the second realisation with the corresponding median and dispersion shown with smaller font. We focus here on the normal-velocity evaporative regime and exclude the high-velocity tail of the distribution.}
\label{discuss1}
\end{figure}

\begin{figure}
\includegraphics[angle=0,width=8.4cm]{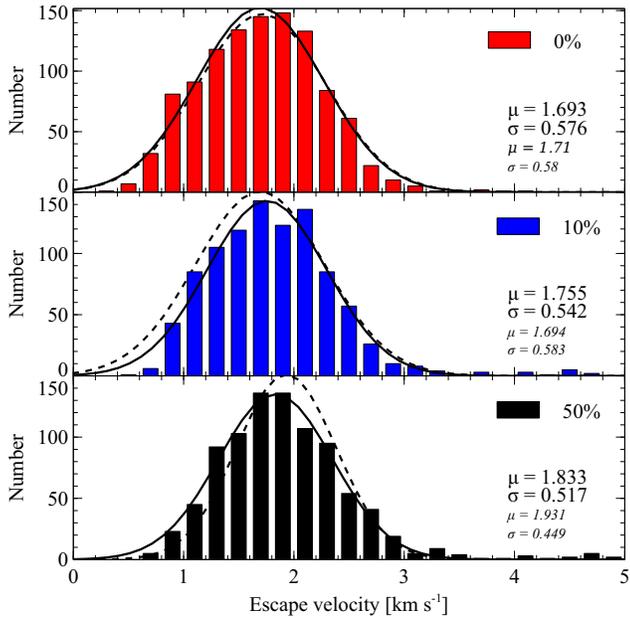}
\caption{As figure \ref{discuss1} but for stars with masses above 0.7 M$_{\odot}$.}
\label{discuss2}
\end{figure}

\begin{figure}
\includegraphics[angle=0,width=8.4cm]{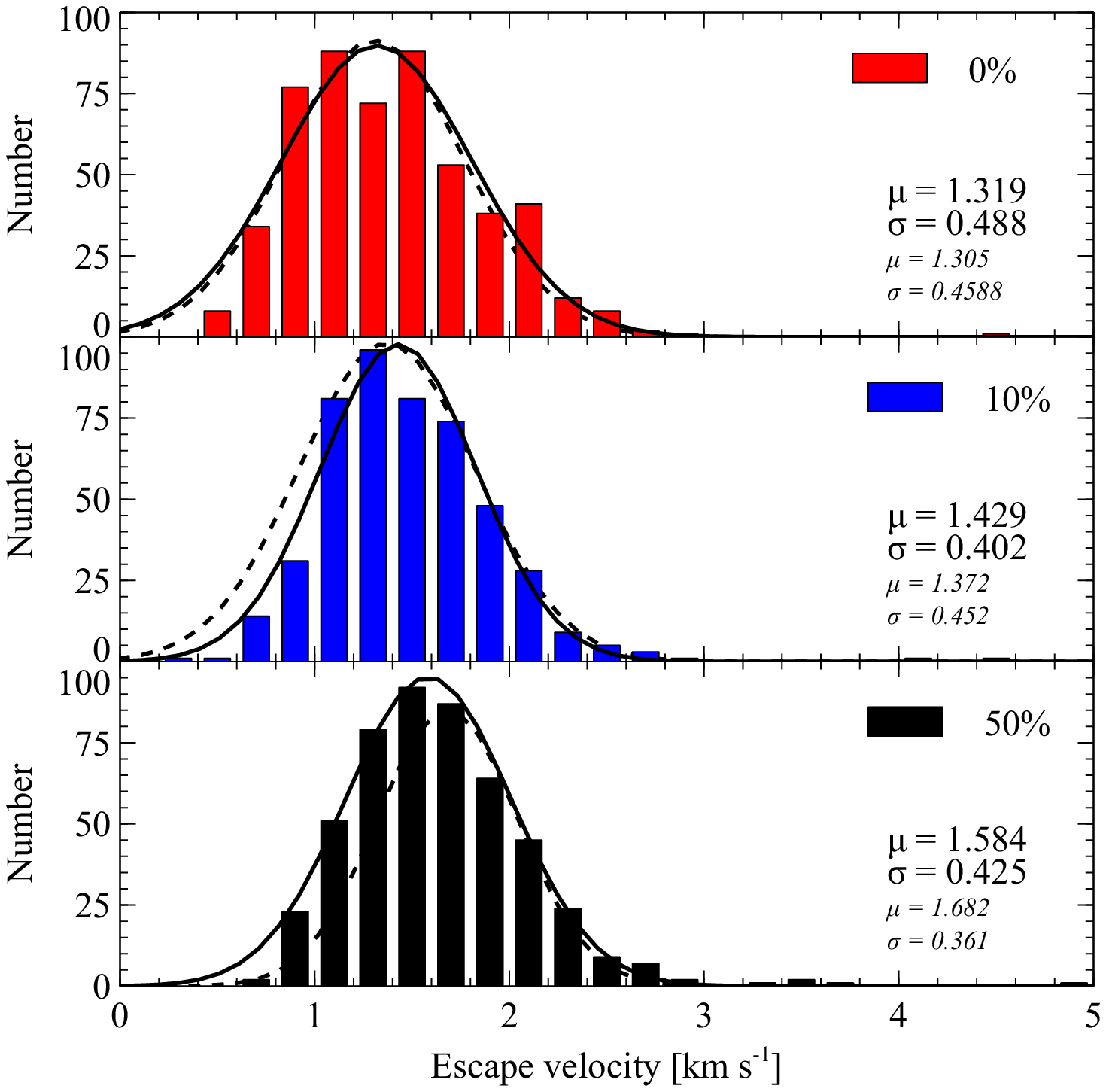}
\caption{As figure \ref{discuss1} but for white dwarfs.}
 \label{discuss3}
\end{figure}

Furthermore, we have analysed the velocity distributions of the normal-velocity stars escaping from our models in Figures \ref{discuss1} to \ref{discuss3} for low-mass main-sequence stars, high-mass main-sequence stars and white dwarfs respectively. In these figures we have calculated the distribution of velocities of the escapers produced by evaporation and low-energy two-body encounters. Even though our normal group of escapers has an upper limit of 20 km s$^{-1}$ we cut the upper velocity to 5 km s$^{-1}$ for the sake of the representation. These histograms were obtained from averaging the information of two realisations. To represent the presence of these two realisations we have included Gaussian fits along with the median and dispersion for each of them. In all cases the median of the distribution increases with increasing primordial binary percentage. The case for white dwarfs is interesting since the environment of the cluster at the moment of escape of these systems is quite different than the environment at the moment of escape of the bulk of the main-sequence stars. These distributions could be affected by the different total mass of the models with different primordial binary percentages, i.e. changes in the escape velocity of the cluster as a whole. We will discuss the implication of the different total mass in Section \ref{DISCUSSION}.


\subsection{Velocity kicks}
\label{KICKS}

As mentioned before, we set a dispersion of $190\, \rm{km}\, \rm{s}^{-1}$ in the distribution of velocities from which the velocity kick is randomly taken at the moment of birth of a neutron star. This value is in agreement with the results predicted by the latest hydrodynamical simulations of natal velocity kicks at the moment of formation of neutron stars, although the magnitude of this threshold is still uncertain (e.g. \citealt{Hills2}; \citealt{Lorimer}; \citealt{Hansen}; \citealt{Hartman}; \citealt{Nordhaus}). Since this uncertainty is present, we analyse different scenarios by running the models {\it{set2a,b,c}} and {\it{set3a,b,c}}. These models differ from our standard set with respect to the dispersion used in the distribution of velocity kicks for white dwarfs, neutrons stars and black holes. {\it{Set2}} introduces a low dispersion (2 km s$^{-1}$) for white dwarfs, while the dispersion for neutron stars and black holes is set to the black hole value in our standard case (6 km s$^{-1}$), representing a low velocity kick. {\it{Set3}} recovers the idea that white dwarfs do not receive any kick at the moment of creation, while the dispersion for neutron stars and black holes is set to the higher neutron star value of our standard case (190 km s$^{-1}$), which means that they will most likely receive a high-velocity kick at the moment of creation.

Besides the obvious difference of not creating hyper-velocity escapers via the {\it{BEM}} mechanism, the models of our {\it{set2}} retain more stars because neutron stars receive only a small kick at the moment of formation. This situation should initiate two competing mechanisms: the increasing probability of a single star to interact with another single star or binary system with the consequent ejection of some of the members interacting, and the increasing difficulty of a star to escape owing to the higher mass of the inner part of the cluster, i.e. a higher escape velocity threshold. Although the total number of neutron stars retained in {\it{set2}} could seem insignificant, the mass within the 10\% Lagrangian radii at 2 Gyr is $\sim 2\%$ larger than in our standard models regardless of the primordial binary fraction. This independence with respect to the binary fraction is expected, since these relaxation-dominated collisional models erase the memory of their initial conditions after a few  relaxation times.

We have analysed the close encounter rate ratio between our standard and {\it{set2}} models by using a simplified version of the relation found by \citet{Davies},

\begin{equation}
\label{erate}
t_{enc} \propto \frac{v_{\sigma}}{n}
\end{equation}

where $n$ is the number density of stars and $v_{\sigma}$ is the velocity dispersion. We use values within the half-mass radius for the comparison. Eq. \ref{erate} gives the close encounter time scale which is inversely proportional to the rate of encounters.

\begin{figure}
\includegraphics[angle=0,width=8.4cm]{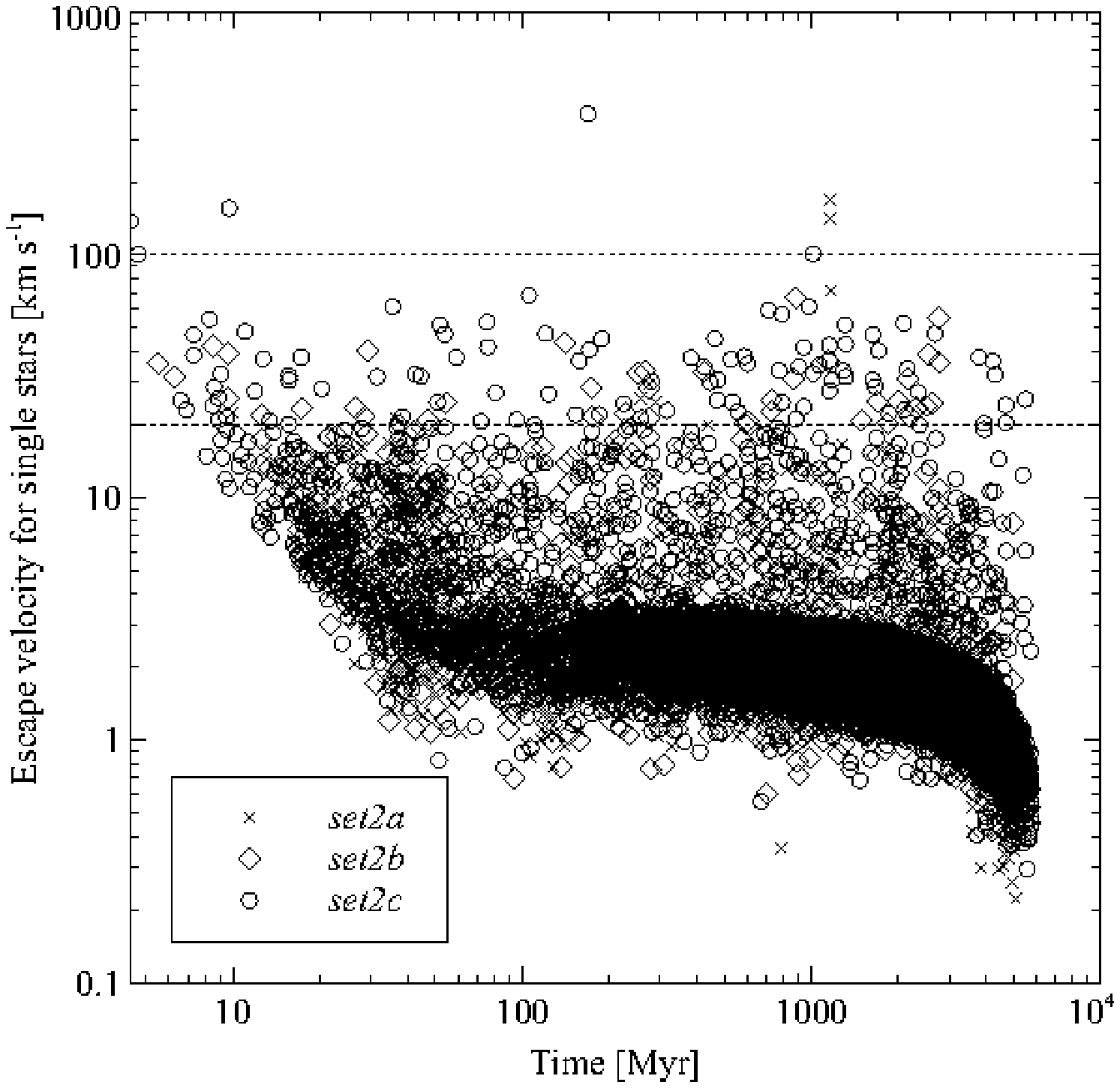}
 \caption{Escape velocities for all the single stars in our {\it{set2}} of models. The escaper with the highest velocity is a O/Ne white dwarf with a velocity of 381.8 km s$^{-1}$ that occurs at 168.6 Myr.}
 \label{velsin-set2}
\end{figure}

The encounter rate is $\sim 50\%$ higher between 0.5-1 Gyr for {\it set2} than for our standard models. This difference results from the higher number of systems, especially in the core regions, indicating that the stars will interact with each other more frequently. As a result, escape velocities within the hyper-velocity range are achieved (see Fig. \ref{velsin-set2} and Table \ref{fig16} for details of these escapers) not only at early times, but also at 1 Gyr. These escapers, which are main-sequence stars, did not occur in our standard set.

Comparing {\it{set1}} and {\it{set2}} escapers within the high-velocity range we find
that the number of escapers was similar for the models with 0\% primordial
binaries (3 stars per model) and 50\% primordial binaries (45 stars per
model). However, the number of high-velocity escapers in the simulations
with 10\% primordial binaries increased to 20 in {\it{set2}} compared to 10 for
the standard model.

\begin{table}
\caption{Hyper-velocity stars escaping from the {\it{set2}} models as depicted in Fig. \ref{velsin-set2}. The columns represent the parent model of the escaper, mass (M$_{\odot}$), velocity (km s$^{-1}$), time (Myr) and evolutionary type at the moment of escape. The last column indicates the mechanism of ejection: disruption of a binary system due to supernova event (binBEM), disruption of a binary system by three (or more) body encounter (binDEM) or pure ejection of a single star due to three-body interaction (sinDEM).}
\label{fig16}
\begin{tabular}{@{}crcrcc}
\hline
Model & Mass & Vel & $T_{esc}$ & Type & Mech\\
\hline
set2a & 0.4 & 168.6 & 1161.6 & Naked He star & binDEM\\
set2a & 1.4 & 140.8 & 1161.6 & Core He Burn & binDEM\\
set2c & 0.2 & 100.6 & 4.7 & MS<0.7 M$_{\odot}$ & sinDEM\\
set2c & 0.2 & 137.5 & 4.4 & MS<0.7 M$_{\odot}$ & binBEM\\
set2c & 1.0 & 101.3 & 1012.4 & MS>0.7 M$_{\odot}$ & binDEM\\
set2c & 1.3 & 381.8 & 168.6 & O/Ne WD & binDEM\\
set2c & 14.5 & 155.9 & 9.7 & MS>0.7 M$_{\odot}$ & binDEM\\
\hline
\end{tabular}
\end{table}

It is important to mention that the higher encounter rates for {\it{set2}} are a consequence of the increased mass in these clusters at a given time which indirectly produces an increase in the density of stars. Since the latter can have a profound impact on the escape rate we further explore the effect in section \ref{density}.

We have explored white dwarf kicks only within our {\it{set2}} models where we included a small kick at the moment of creation of these stars. In Fig. \ref{denserall} we can see that the velocity distribution of the bulk of the white dwarfs escaping is similar between the models regardless of the kick assumption. We could not find any noticeable difference between our standard models and those of {\it{set2 }} when we searched for white dwarf escapers within the high-velocity escape range, in particular these were once again only produced in the model with 50\% binaries.

 For completeness, in our {\it{set3}} models we explored the case where neutron stars and black holes are born by the same mechanism, i.e. the same velocity kick distribution with a dispersion of $190\, \rm{km}\, \rm{s}^{-1}$ is imprinted on both type of remnants. In this final case, we returned to the assumption that white dwarfs do not receive any kick.
This last set of models ({\it{set3}}) has quite similar behaviour to our standard models with respect to the number and velocity of the escaping systems. Of particular interest is that we found in our results two main-sequence stars escaping with hyper-velocities. Combined with the results of {\it{set2}} this suggests that the dynamics at the moment of creation of stellar black holes and  neutron stars play a crucial role in allowing escapers to reach hyper-velocities, or at least collaborating in stirring stellar clusters to set the conditions for hyper-velocity escapers to occur.

Since black holes in clusters tend to sink to the inner part of the cores and even create binary systems with other massive stars (or a black hole companion), they produce highly noisy cores in stellar clusters because they kick other less massive systems from this inner region. In the {\it{set3}} models neutron stars and black holes are not retained. Without the energy input of these objects participating in two-body encounters the modelled stellar systems grow, i.e. they show an increase in the half-mass radius, in a smoother way than in the {\it{set1}} models. In an opposite way, the models corresponding to {\it{set2}} have the noisiest cores because they retain more black holes and neutron stars.

\subsection{Initial density of the stellar clusters}
\label{density}
We next further explore the role that stellar density plays in the escape rate of stars from their parent cluster. In order to do this analysis we have reduced the initial sizes of our clusters by half but maintained the escape criteria of 100 pc and ensured the clusters are initially in virial equilibria (for analysis without initial virial equilibria we refer the reader to \citealt{Allison2}; \citealt{Enrico} and references therein).

\begin{figure}
\includegraphics[angle=0,width=8.4cm]{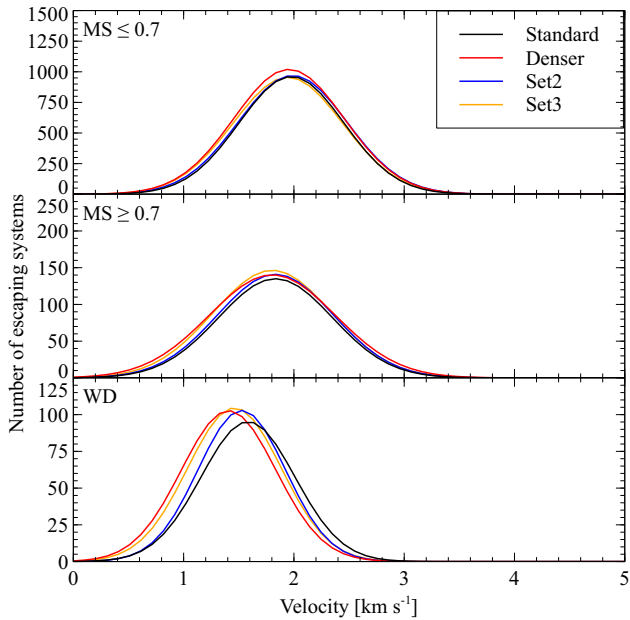}
 \caption{Gaussian fits for single stars escaping in all the models with 50\% of primordial binaries (black for our standard model, red for the denser ones, blue for {\it{set2}} and yellow for {\it{set3}}). The different panels show the number of systems escaping as a function of their velocity (upper panel: main-sequence stars with masses below 0.7 M$_{\odot}$, middle panel: main-sequence stars with masses above 0.7 M$_{\odot}$, lower panel: white dwarfs.}
 \label{denserall}
\end{figure}

To analyse the effect of having a different initial concentration for our clusters we have compared in Fig. \ref{denserall} the Gaussian fits to the distribution of single stars escaping according to their velocity. These fits are similar to those made in Figs. \ref{discuss1} to \ref{discuss3} but now we do not show the histograms. 

Even with the velocity distribution of the bulk of the stars escaping through evaporation processes being similar for all the sets of models that we have run (see Fig. \ref{denserall}), in all cases we found that the denser models retain less systems over time than their less dense counterparts. At first it might be assumed that this means that denser models have a higher encounter rate of stars, producing a larger number of escapers. However, we found that this was not the case in practice. This can be explained by recalling Eq. \ref{erate}, where the encounter rate is not only a function of the average density but also the velocity dispersion. In our denser models the velocity dispersion of stars is higher than the other models, making the encounter rate of the denser models similar to our standard simulations. Instead it is a consequence of the shorter half-mass relaxation time for the denser clusters (see Table \ref{THE MODELS}), meaning that evaporation from cumulative weak two-body encounters proceeds at a greater rate.

\section{DISCUSSION}
\label{DISCUSSION}

For the first time we have explored the effect that two key factors will have on the origin and velocity of stellar systems escaping from realistic stellar clusters: the velocity kick distribution given to neutron stars and black holes at birth and the primordial binary fraction, considered consistently together.

In this work we have included not only a realistic mass spectrum for single stars and primordial binaries but also stellar evolution for all stars. We have evolved our clusters until they were almost completely disrupted, a situation that occurred after 4 Gyr. Evolving the modelled clusters up to this stage allowed us to follow their dynamical and stellar evolution history in full, thus expanding upon the work done by \citet{Allison} and \citet{Perets}. We included supernova kicks for the events that create black holes and neutron stars in our standard models. This implementation allowed us to understand the impact of supernova events on the origin of early escapers as a result (in most cases) of the {\it{BEM}} mechanism. In our standard models with strong neutron stars kicks, after the initial violent phase of stellar evolution, i.e. after all the supernova events went off (which occurs roughly before the first 100 Myr), the ejection owing to dynamical interactions ({\it{DEM}} mechanism) became dominant. Even with the presence of runaway stars with high-velocities resulting from close two-body interactions, we found that without strong velocity kicks during the supernova events we can not produce what we called hyper-velocity neutron stars or black holes as a product of the {\it{DEM}} mechanism (see Fig. \ref{velsin-set2}). It seems imperative to have disruption events of binary systems owing to supernovae events to explain hyper-velocity X-ray sources runaways, i.e. escapers with velocities $\ge 1\,000\, \rm{km}\, \rm{s}^{-1}$, from stellar clusters and this is in agreement with \citet{Napiwotzki}. 

The velocity kicks assumed for the later stellar evolution stages will affect the balance between the escapers that are produced via the {\it{BEM}} mechanism \citep{Blaauw} and the {\it{DEM}} mechanism \citep{Poveda}. While including velocity kicks at the commonly assumed levels in our standard models obviously enhances the number of early high- and hyper-velocity escapers, drastically decreasing the kick strength succeeded in creating main-sequence stars escaping with velocities higher than $100\, \rm{km}\, \rm{s}^{-1}$ (up to $170\, \rm{km}\, \rm{s}^{-1}$: see Fig. \ref{velsin-set2}) from {\it{DEM}}, although failed at creating hyper-velocity neutron stars. Obviously we are limited in our ability to decrease neutron stars kicks while maintaining realistic models \citep{Nordhaus}, but the models run within our {\it{set2}} gave us the insight that a higher inner stellar density is needed, i.e. having more stellar systems (white dwarfs /neutron stars/black holes) as a result of not giving them a velocity kick at the moment of creation, while keeping the cluster cold in order to boost two-body interactions and in this way reduce the encounter time if we want main-sequence hyper-velocity escapers from the {\it{DEM}} mechanism.

	We have compared the Gaussian fits in the normal-velocity range, i.e. stars dominated by evaporation, for all our models with 50\% of primordial binaries in Fig. \ref{denserall}. We can see from this figure that regardless of the velocity kicks assumed for neutron stars and black holes or the initial density of the clusters,  main-sequence stars escaping through evaporation processes are affected in a similar way. More notorious is the case of the white dwarfs in the non-standard models. The environment of the cluster at the moment of the onset of white dwarfs will determine the velocity ranges for the escape of these systems. In general, the non-standard models have less mass at the moment of creation of white dwarfs -- the denser models lose mass at a faster rate than the other models, {\it{set2}} loses mass more rapidly than the standard models since the number of two-body interactions is higher and {\it{set3}} lost all of its neutron stars and black holes -- which means a lower velocity is required in order to escape. This higher evaporation rate for the denser models causes stars to reach the escape velocity earlier in the history of the cluster. However, the distribution of escape velocities through evaporation is not affected as can be seen in Fig.\ref{denserall}.

The encounter rate is directly affected by the primordial binary fraction. Having a higher number of primordial binary systems will enhance the number of two-body interactions within the stellar clusters owing to the larger gravitational cross section of binary systems, as studied by \citet{HeggieHut}. This mean that stars in stellar clusters with a high number of binaries will interact more frequently and in this way change their velocities on a shorter timescale, producing on average an enhancement of the number of escaping systems.

When we look at the distributions of escape velocities (Figs. \ref{discuss1}-\ref{discuss3})
we find a general trend for the mean of the Gaussian distributions
to increase with the primordial binary fraction of the stellar system.
Differences of up to 13\% for high-mass main-sequence stars and
29\% for white dwarfs are found when comparing simulations with
0\% and 50\% of binaries.
Before attributing this result to the binary content we must first
determine if any other factors are playing a role, particularly
any relation with the escape velocity of the model clusters.
The escape velocity evolves in a similar way for each of our models,
steadily decreasing in time as the cluster mass decreases
(dropping by $\sim80$\% between 0 to 4 Gyr).
However, our decision to keep the total number of systems
(stars and binaries) constant across our initial models while
varying the primordial binary fraction means they have different
initial mass.
This translates to different escape velocities.
Comparing the escape velocities at the half-mass radius between
models with 0\% and 50\% primordial binaries we find a difference
of $\sim20$\% (focussing on conditions at 0 and 2 Gyr).
The cluster with 0\% primordial binaries has the lower escape velocity
at all times.
This difference is similar to that found between the mean velocities
of the respective Gaussian velocity distributions.
To check on the effect we have performed additional models that
keep the same initial total mass for varying binary content
(0\% and 50\%).
These models have the same initial escape velocity, by design,
and the differences grow to at most 10\% at late times in the
evolution (4\% on average).
In this case the model with 50\% binaries has the lower value.
Even so, the velocity distribution outcomes that we find are
similar to those of the standard models and the difference in the
mean of the escape velocity distributions with binary fraction is
preserved.
This suggests that the binary fraction is playing a definite role,
presumably through an increased close encounter rate, and that
this is prevailing over the effect of the gravitational attraction
of the cluster as a whole.

Cases of standard mass transfer and common-envelope episodes in binary systems are frequent in clusters with significant binary populations, whether these be primordial or formed dynamically through exchanges \citep{Heggie75}. All of these mass transfer processes can pollute the original chemical composition of these stars, rejuvenating them when no (or light) processed elements are accreted (e.g. accretion of H or He) or aging them when outer stellar layers are stripped (e.g. removing H which is the main fuel of stars during the main-sequence) exposing their inner structure. We found that a small number, e.g. 30 single stars in each simulation, of escapers were polluted and within them only 20\% were main-sequence or giants stars. 
Anomalous stars may also be produced by mergers -- either via common-envelope or the occasional direct collision -- such as the cases highlighted in Table \ref{escap}.
Our findings will not affect results such as \citet{Martell} but is something to factor in when searches of the origin of polluted stars in the halo of the Galaxy are performed.

In this work we have limited our primordial binary systems to have periods less than approximately 10$^{6}$ d. Wider binaries which are weakly bound will most likely be disrupted by early two-body encounters. So effectively we are pre-empting this disruption and including the soft binary components as single stars from the beginning. If instead we extended our primordial binary population to include periods up to 10$^{10}$ d then our models with 50\% binaries would represent an $\sim$ 70\% binary population which is in line with the findings of \cite{King}. To investigate the effect of a high binary frequency further we have evolved an additional preliminary model with 10\,000 systems and 70\% primordial binaries (once again capped at 10$^{6}$ d so that effectively this represents 100\%  binaries when compared with \citealt{King}). Looking at the number of high-velocity escapers between 200 Myr and 2 Gyr -- to avoid the effects of supernova-induced velocity kicks in the early stages and when the evolution becomes dominated by the external field of the Galaxy in the latter stages -- we find that the number increases from 18$\pm$3 in our 50\% models (averaged over the two realisations of {\it{set1c}}) to 26 for the preliminary 70\% model. This confirms that the primordial binary fraction has a noticeable impact on the frequency of high-velocity escapers from an open cluster. Thus models with such high binary fractions demand further exploration. Also of importance will be the choices made for the distributions of orbital parameters within our primordial binary population. Hard binary systems govern the dynamical behaviour of stellar clusters since they contain a high proportion of the energy budget of the stellar systems. We have explored just one distribution of orbital parameters, which meant that 50\% of the primordial binaries were hard binaries, but the energy spectrum and number distribution of these binary systems could be different \citep{Tanikawa}. Having a different distribution of primordial binaries will produce for instance core collapse at different times and with a different depth, leading to different stellar encounter rates between different models.

For our modelled clusters we have chosen nearly circular orbits, for which the external potential exerted by the Galaxy will be nearly constant. Elliptical orbits affect the rate at which stars escape since the clusters would move in a changing tidal field. \citet{Webb} analysed the influence of the tidal field on the evaporation rate of globular clusters with different elliptic orbits. In their study the key feature to infer the influence of the potential field of the Galaxy on the size of the clusters was that the period of the orbit was shorter than the relaxation time of the clusters. 
For our open clusters models where the relaxation time is shorter than the orbital period, combined with the nearly circular orbit, the role played by the Galaxy is no more than providing a limit to the attraction of the cluster to its members. A stronger effect would be evident if the modelled clusters had shorter orbital periods and/or travelled in a changing external potential (e.g. they were placed on non-circular orbits closer to the Galactic centre). Another important effect that can be relevant is if the orbits had higher inclinations (e.g. disc shocking) and this could be independent of the relaxation time of the cluster itself. Disc shocking could potentially be a key factor that affects the dissolution rate and sizes of the modelled clusters.

We have chosen a continuous IMF for our modelled clusters, but stochastical sampling in the IMF (\citealt{Eldridge} and references therein) could play a role at the moment of formation of these low mass stellar systems. This sampling could affect the distribution of masses within our cluster and change the nature of the stars that will escape from these stellar clusters, particularly through fluctuations of the small number of massive stars.

Also, it is now accepted that the metallicity can change the overall evolution of a stellar cluster (e.g. \citealt{Hurley04}; \citealt{Schulman}), with the disruption timescale and the analysis of the escapers potentially affected by changes in the metallicity. We will try to address the effect of this parameter in the near future.

As we expand our studies, IMF, metallicity and binary population variations will be considered, as will different N. This will be part of the TraCD ({\bf{Tra}}cing {\bf{C}}lusters {\bf{D}}ebris) program (Moyano Loyola, Gibson, Flynn, Hurley in prep.) with the main aim of tracing back the dissolution history of stellar clusters within the potential of our Galaxy. Following post-escape velocities and trajectories in more detail using new models that include diffusion and churning (Flynn, personal communication) will also allow us to map the high-velocity stars that reach the halo with greater certainty.

\section{CONCLUSIONS}

Preparing the field to harvest all the information that GAIA\footnote[1]{\href{http://sci.esa.int/science-e/www/area/index.cfm?fareaid=26}{ESA Science and Technology: GAIA}} and HERMES\footnote[2]{\href{http://www.aao.gov.au/HERMES/}{AAO: The HERMES project}} will provide has been a primary objective of a large
number of astronomers during the last decade, not only from an observational point of view but theoretical and computational
as well.

Noting that many stars in the Galaxy originate from star clusters
we were motivated to determine if the binary content of a star
cluster can leave a noticeable imprint on the velocity distribution
of the stars that escape into the Galaxy.
Comparing Gaussian fits to the velocity distributions of stars escaping
from models with 0\% and 50\% primordial binaries, we find that the mean
of the distribution increases slightly with increasing binary fraction.
Differences of up to 13\% for main-sequence star and 29\% for white
dwarf population distributions are found, primarily attributed to
an enhanced two-body interaction rate arising from the greater
number of binaries but also with a secondary effect related to
differing escape velocities of the models.
However, overall it was evident that the percentage of binaries
in a star cluster population does not produce significant variations
in the velocities of the stars that escape in the normal-velocity
range (less than 20 km s$^{-1}$), which represents 99\% of the escapers. For binaries we did not find any significant correlation between the
escape velocity of a binary and its orbital properties.

Looking at distributions by stellar type we do find evidence of trends
with primordial binary fraction, particularly in the high-velocity
range (20-100 km s$^{-1}$).
For main-sequence stars the number of high-velocity escapers increased
markedly as we moved from models with 0\% primordial binaries to 10\%
and then 50\%.
Simarly, the incidence of high-velocity white dwarf escapers increased
by a factor of four for 50\% binaries compared to 0\%, although the result
is subject to low-number statistics in this case.

We made a population census of our standard models where we quantified the number of stars in the clusters at a given Gyr (0, 1, 2, 3 and 4 Gyrs) and the stars that escape in the Gyr prior to the census time. As expected, owing to the velocity kicks assumed for neutron stars and black holes, these remnants leave the clusters in the first Gyr, while the number of white dwarf escapers slowly increases as time passes. It is important to mention that the environment of the cluster is quite different at the moment of escape of these respective systems.
The binary content of the escapers was 8\% from clusters that started
with 10\% binaries and 27\% for clusters that started with 50\% binaries.

Different simulations have been tried in the past to account for the origin of runaway stars. The initial conditions and settings for the physical effects studied have included simulations without primordial binaries \citep{Fujii} and simulations without stellar evolution (\citealt{Allison}; \citealt{Perets}), among the latest examples, and all of them have been successful in explaining some aspects of the nature of the runaway stars from stellar clusters.

In our work, we could determine that the assumptions made at the moment of the creation of the heavy remnants, i.e. white dwarfs, neutron stars and black holes, play a key role in the velocities that the stars escaping from the stellar system can achieve. Using values of velocity kicks for supernovae events in agreement with \citet{Nordhaus} we can account for hyper-velocity black holes and neutron stars, which will be possible candidates for runaway X-ray sources. Runaway stars with velocities higher than 1\,000 km s$^{-1}$ may still need a different mechanism, such as dynamical ejection resulting from a two-body interaction with the central supermassive black hole of the Galaxy (\citealt{Bromley} and references therein).

For our standard models, with neutron star and black hole velocity
kicks chosen at random from distributions with dispersions of 190
and 6 km s$^{-1}$ respectively, we did not find any hyper-velocity
main-sequence or giant escapers. However, for models where we imposed lower velocity kicks for neutron
stars, thus keeping more stars within the model and enhancing the
two-body encounter rate, hyper-velocity main-sequence stars escaping
through the dynamical ejection mechanism were produced.
Conversely, this also occurred in models where both the black hole and
neutron star velocity kick dispersions were set high (190 km s$^{-1}$),
suggesting that the high incidence of energetic supernova remnant
escapers stirred the inner regions of the cluster and led to energetic
two-body encounters.
We also looked at models with an increased initial stellar density
which would be expected to have an increased number of two-body
encounters but found no hyper-velocity main-sequence stars.
A further analysis of the interplay between the initial density,
number of systems, initial mass function and primordial binary fraction
will be explored in future models.

Analysing our results we can infer that dissolving stellar clusters such as those that we have modelled can populate the Galactic halo with giants stars for which the progenitors were stars of up to $\sim$ 2.4 M$_{\odot}$. This limit will be affected by the ejection velocities of the escapers. Boosting the mechanisms that cause main-sequence stars to escape, i.e. increasing the percentage of binary systems which will increase the number of energetic close encounters, would produce escapers with velocities that will allow more massive stars to reach higher escape velocities and in this way reach the Galactic halo prior to evolving into white dwarfs.

\begin{table*}
\begin{minipage}{\textwidth}
\caption{Population census for single stars (upper) and binaries (lower) in the standard models. On the left is the number of stars of different evolutionary stages present in the clusters at a given time. The ``giants'' column comprises stars passing the Hertzsprung gap, stars on the giant branch, stars burning helium in their cores, early AGB stars and TPAGBs (i.e. k-type 2 to 6: see \citealt{Hurley1}). The white dwarf column comprises He WDs, C/O WDs and O/Ne WDs. The NS/BH column is the sum of neutron stars and black holes. On the right is shown the number of stars that have escaped within each gigayear interval, i.e. in the first, second, third and fourth gigayear. Each entry in this table is the average between two realisations for the standard models. By analysing the first two main-sequence columns on the left it is evident how the systems evolve to retain more ``heavy'' stars which is a consequence of energy equipartition. The columns for the population census for binary stars are showing the same information as for single stars, with the difference that the numbers in each column represent the numbers of stars within binaries. Each entry for the binary census is the average of two realisations, so adding all the values for a given row could return an odd number in some cases.}
\label{census}
\begin{tabular}{@{}lrrrrrrrrrrrr}
\hline
&AGE &  &  &  & & & \vline & & & & & \\
&(Gyr) & MS$\le$0.7$M_{\odot}$ & MS$\ge$0.7$M_{\odot}$ & giants & WDs & NS/BH&\vline & MS$\le$0.7$M_{\odot}$ & MS$\ge$0.7$M_{\odot}$ & giants & WDs & NS/BH\\
0\% & 0 & 8155 & 1845 & 0 & 0 & 0&\vline & - & - & - & - & - \\
    & 1 & 5649 & 1434 & 75 & 259 & 0&\vline & 2497 & 263 & 5 & 13 & 69\\
    & 2 & 2894 & 1137 & 48 & 342 & 0&\vline & 2761 & 243 & 3 & 69 & 0\\
    & 3 & 1005 & 784 & 47 & 310 & 0&\vline & 1887 & 225 & 2 & 122 & 0\\
    & 4 & 205 & 393 & 35 & 176 & 0&\vline & 804 & 207 & 10 & 175 & 0\\
\hline
10\% & 0 & 7358 & 1638 & 0 & 0 & 0&\vline & - & - & - & - & -\\
     & 1 & 5293 & 1422 & 70 & 262 & 1&\vline & 2427 & 254 & 6 & 17 & 66\\
     & 2 & 2794 & 1156 & 47 & 363 & 0&\vline & 2545 & 247 & 5 & 59 & 0\\
     & 3 & 1036 & 801 & 38 & 305 & 0&\vline & 1774 & 219 & 7 & 147 & 0\\
     & 4 & 241 & 421 & 26 & 171 & 0&\vline & 806 & 209 & 4 & 183 & 0\\
\hline
50\% & 0 & 4069 & 930 & 0 & 0 & 0&\vline & - & - & - & - & -\\
     & 1 & 4092 & 1282 & 77 & 306 & 1&\vline & 1780 & 302 & 4 & 25 & 82\\
     & 2 & 2389 & 1107 & 48 & 421 & 0&\vline & 1908 & 246 & 4 & 74 & 0\\
     & 3 & 1168 & 832 & 46 & 387 & 1&\vline & 1384 & 204 & 9 & 135 & 0\\
     & 4 & 428 & 518 & 35 & 274 & 1&\vline & 772 & 175 & 6 & 171 & 0\\
\hline
\hline
0\% & 0 & 0 & 0 & 0 & 0 & 0&\vline & - & - & - & - & - \\
    & 1 & 0 & 4 & 0 & 3 & 0&\vline & 2 & 1 & 0 & 1 & 1\\
    & 2 & 0 & 2 & 2 & 4 & 0&\vline & 1 & 0 & 0 & 2 & 0\\
    & 3 & 0 & 2 & 0 & 6 & 0&\vline & 1 & 1 & 1 & 0 & 0\\
    & 4 & 0 & 2 & 0 & 3 & 0&\vline & 2 & 1 & 0 & 1 & 0\\
\hline
10\% & 0 & 1588 & 416 & 0 & 0 & 0&\vline & - & - & - & - & -\\
     & 1 & 879 & 175 & 6 & 4 & 0&\vline & 319 & 7 & 1 & 18 & 1\\
     & 2 & 514 & 109 & 2 & 8 & 1&\vline & 333 & 8 & 1 & 17 & 0\\
     & 3 & 271 & 69 & 3 & 8 & 0&\vline & 232 & 5 & 0 & 13 & 1\\
     & 4 & 97 & 39 & 1 & 6 & 0&\vline & 176 & 3 & 1 & 9 & 0\\
\hline
50\% & 0 & 7989 & 2012 & 0 & 0 & 0&\vline & - & - & - & - & -\\
     & 1 & 4529 & 851 & 11 & 49 & 2&\vline & 1596 & 40 & 1 & 83 & 2\\
     & 2 & 2850 & 572 & 9 & 71 & 2&\vline & 1661 & 24 & 1 & 90 & 0\\
     & 3 & 1647 & 409 & 8 & 75 & 0&\vline & 1290 & 18 & 0 & 70 & 1\\
     & 4 & 788 & 266 & 11 & 62 & 0&\vline & 868 & 18 & 0 & 45 & 0\\
\hline

\end{tabular}  
\end{minipage}
\end{table*}

\section*{ACKNOWLEDGEMENTS}

We thank the anonymous referee for his suggestions that helped to improved this work. This work was performed on the gSTAR national facility at Swinburne University of Technology. gSTAR is funded by Swinburne and the Australian Government's Education Investment Fund. GML wants also to thank Swinburne University of Technology for the SUPRA scholarship.

\bibliography{biblio}
\bibliographystyle{mn2e}

\appendix
\section{State of Our Modelled Clusters}

Here we show the X-Y projection (in parsecs) of our clusters at 0, 2, 3 and 4 Gyr. Since we set our escape criteria to 100 pc we can see in the figures the tidal stream of stars that are leaving the clusters as time pass. The x-axis points toward the Galactic anticentre and the z-axis to the Galactic North Pole. Previous studies \citep{Royer} had mentioned a predominance of escapers towards the U-velocity, i.e. Galactic anticentre. This will be a matter of future study as we will explore different orbits for our modelled clusters.

\begin{figure*}
\centering
\begin{tabular}{cc}
\includegraphics[width=0.45\textwidth]{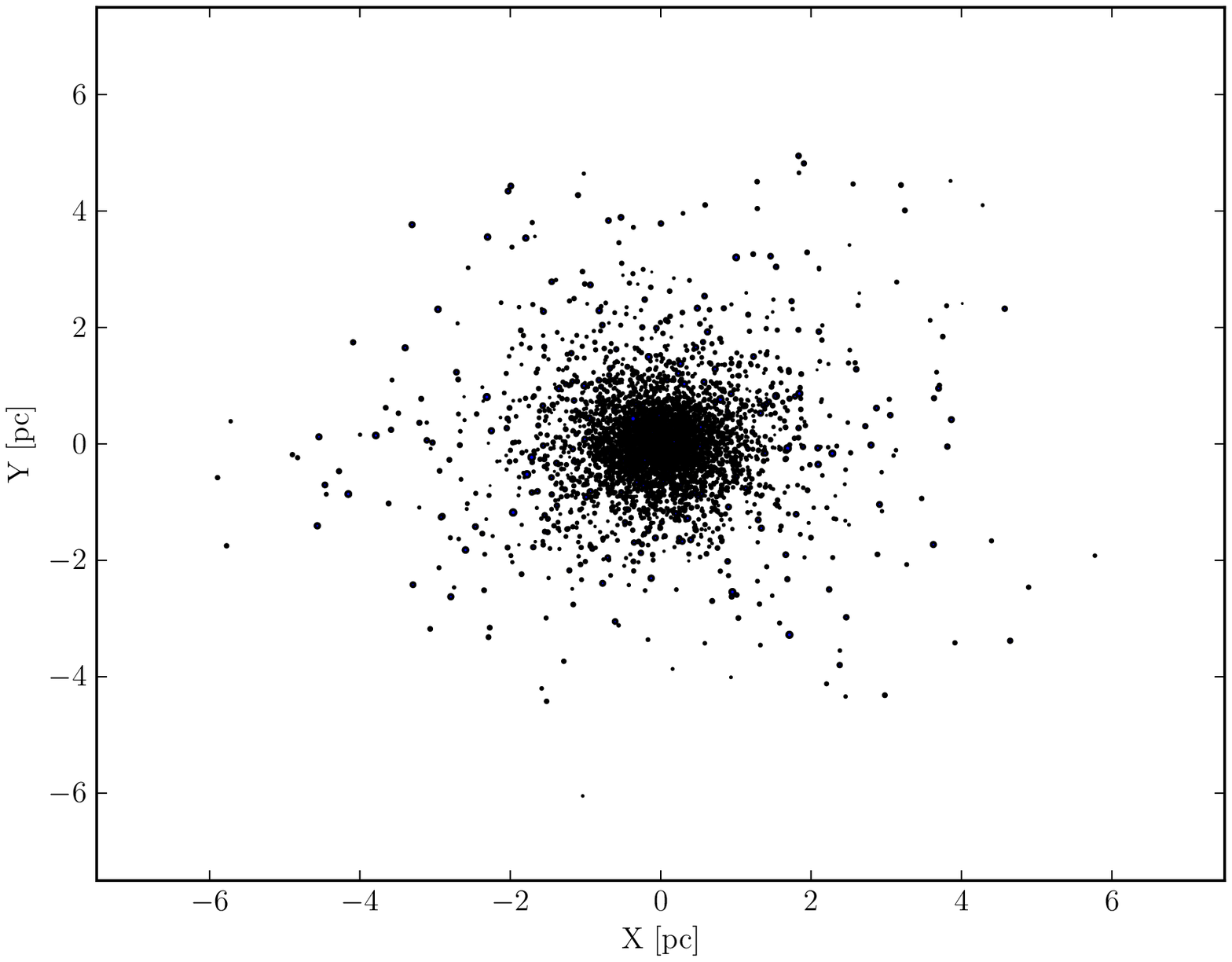} &
\includegraphics[width=0.45\textwidth]{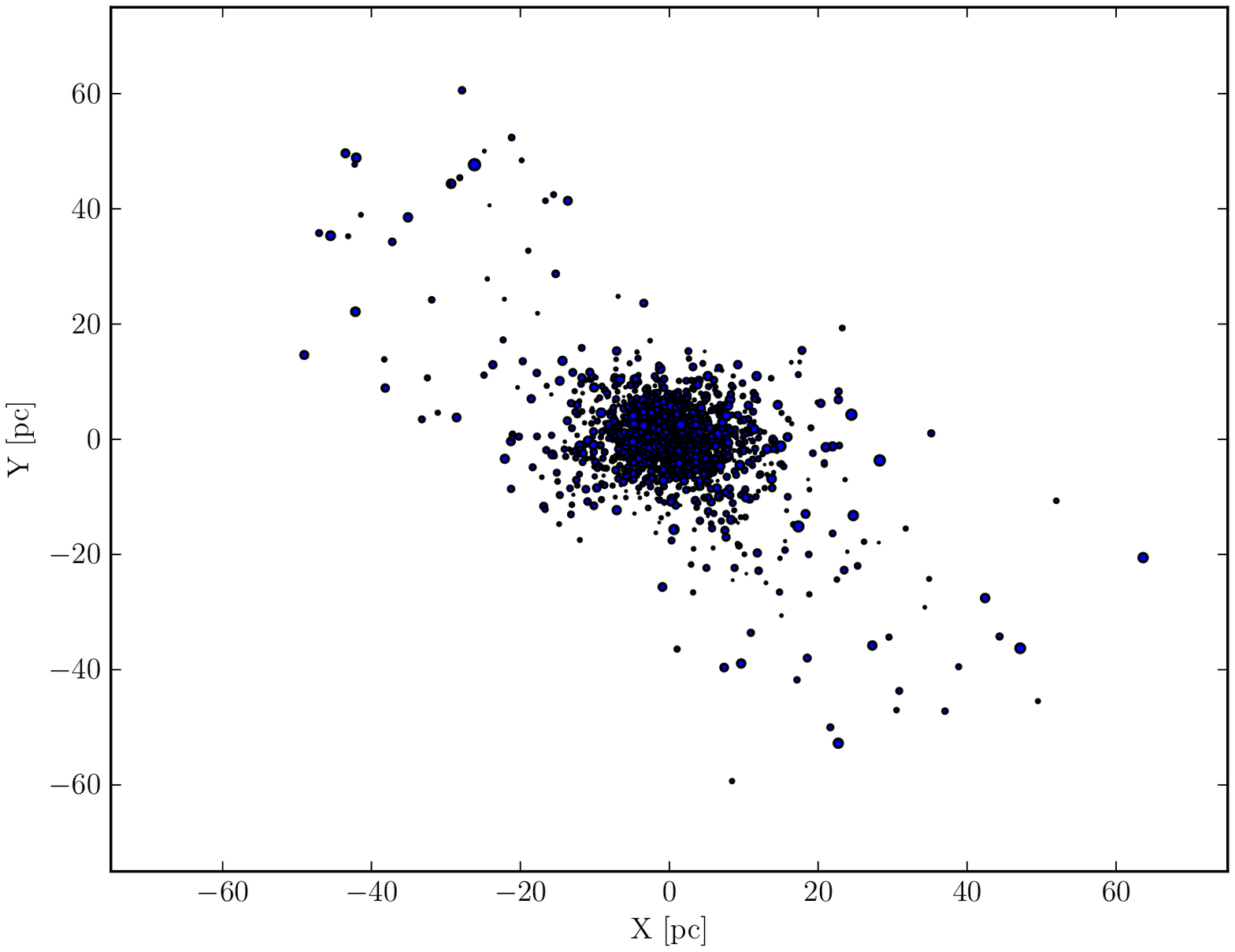} \\
\includegraphics[width=0.45\textwidth]{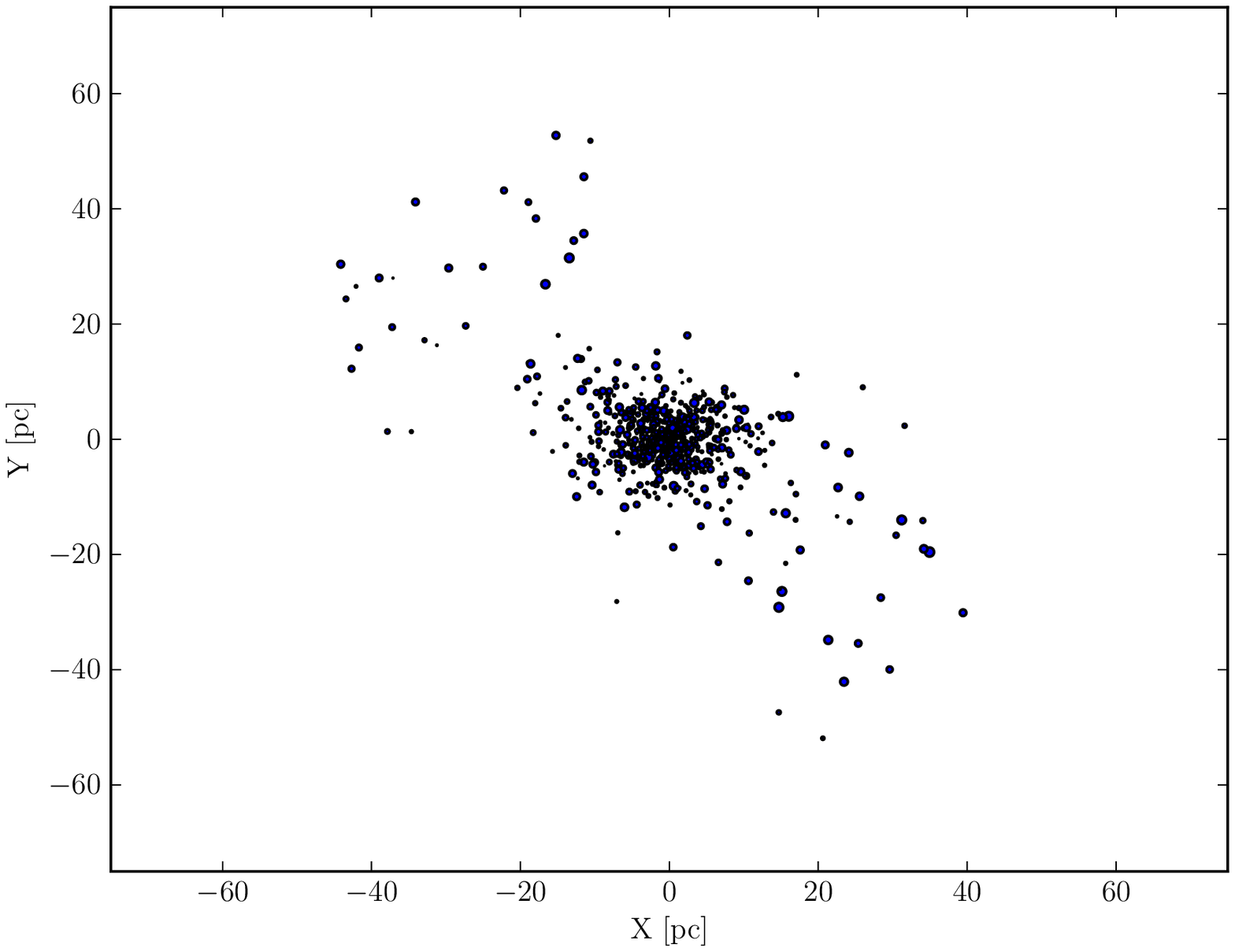} &
\includegraphics[width=0.45\textwidth]{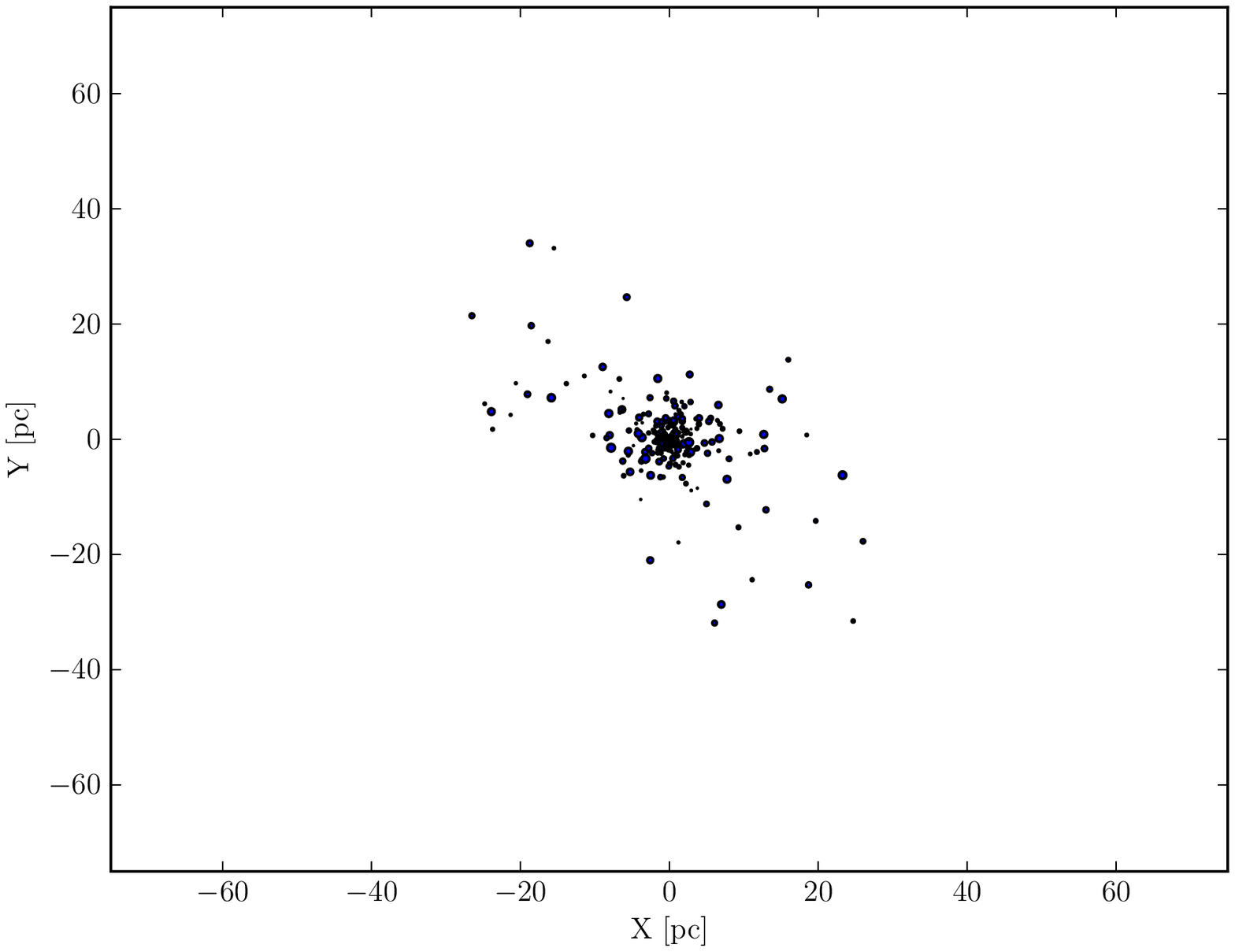}
\end{tabular}
\caption{Two-dimensional projection in parsecs (x-y plane of the Galaxy) of our {\it{set1a}} model at 0 Gyr (top-left panel), 2 Gyr (top-righ panel), 3 Gyr (bottom-left panel) and 4 Gyr (bottom-right panel). The sizes of the points in the figure correlate with the z-distances (the larger the points, the closer to the reader). In the figure it can be seen the tidal streams of escaping stars.}
\label{appen1}
\end{figure*}

\begin{figure*}
\centering
\begin{tabular}{cc}
\includegraphics[width=0.45\textwidth]{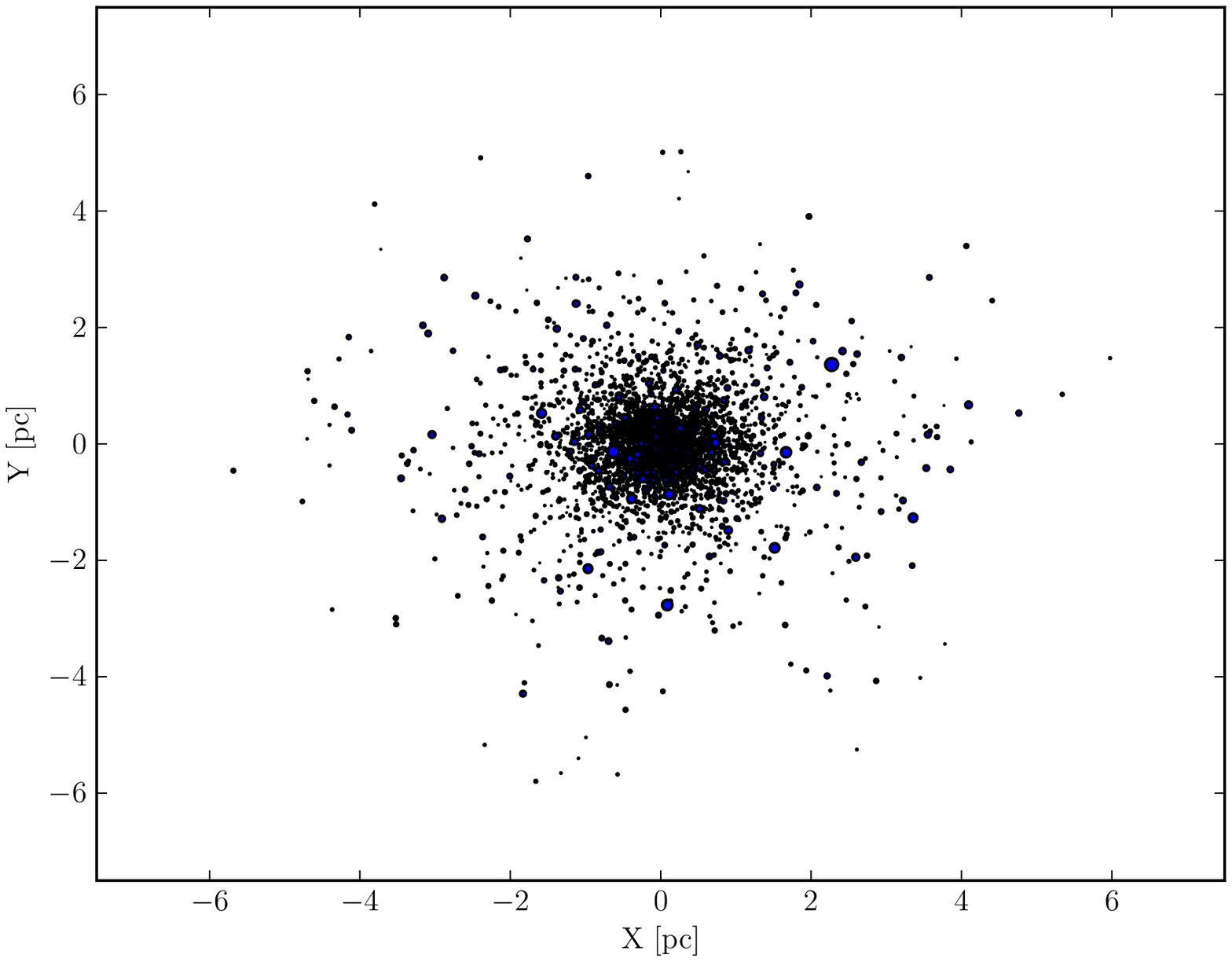} &
\includegraphics[width=0.45\textwidth]{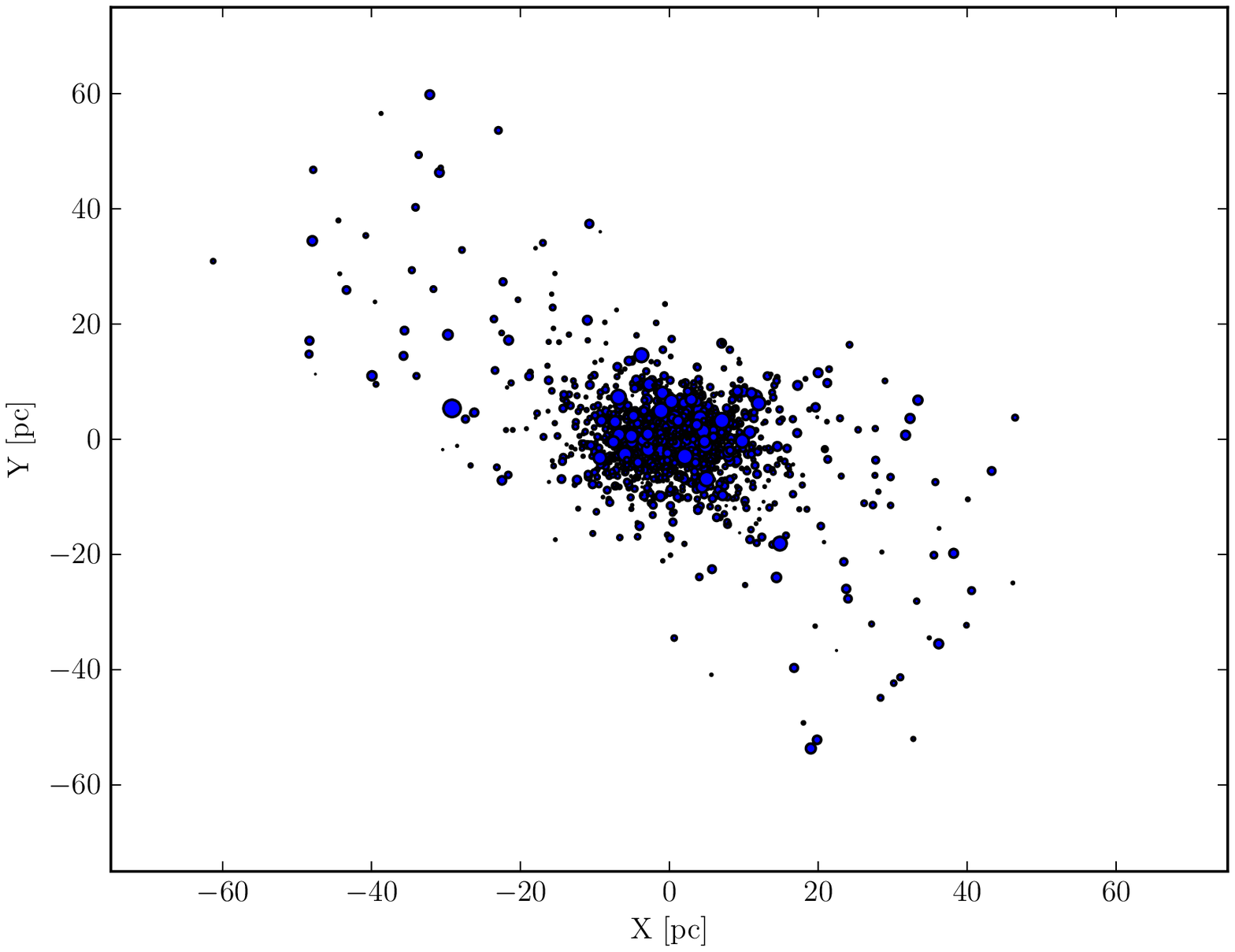} \\
\includegraphics[width=0.45\textwidth]{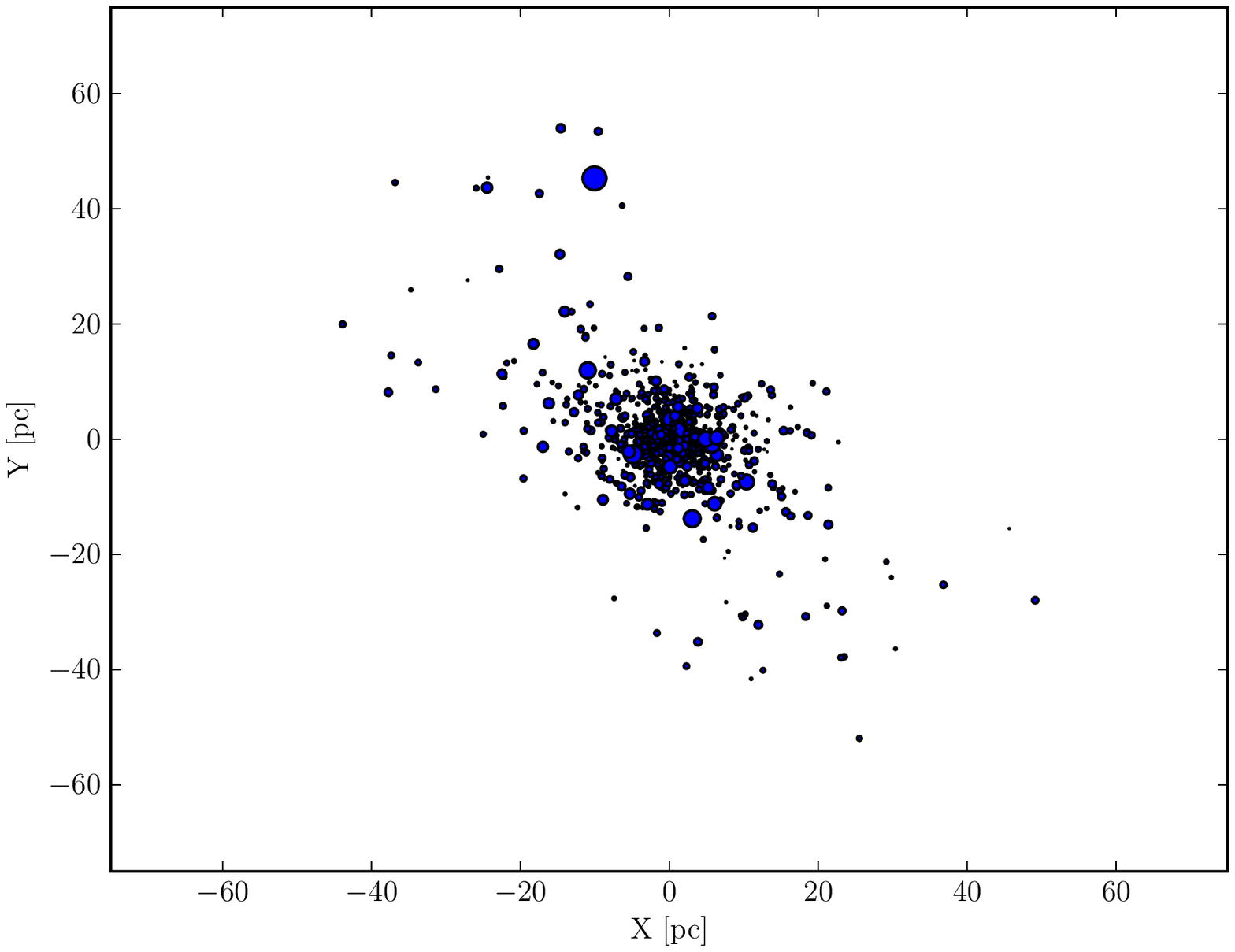} &
\includegraphics[width=0.45\textwidth]{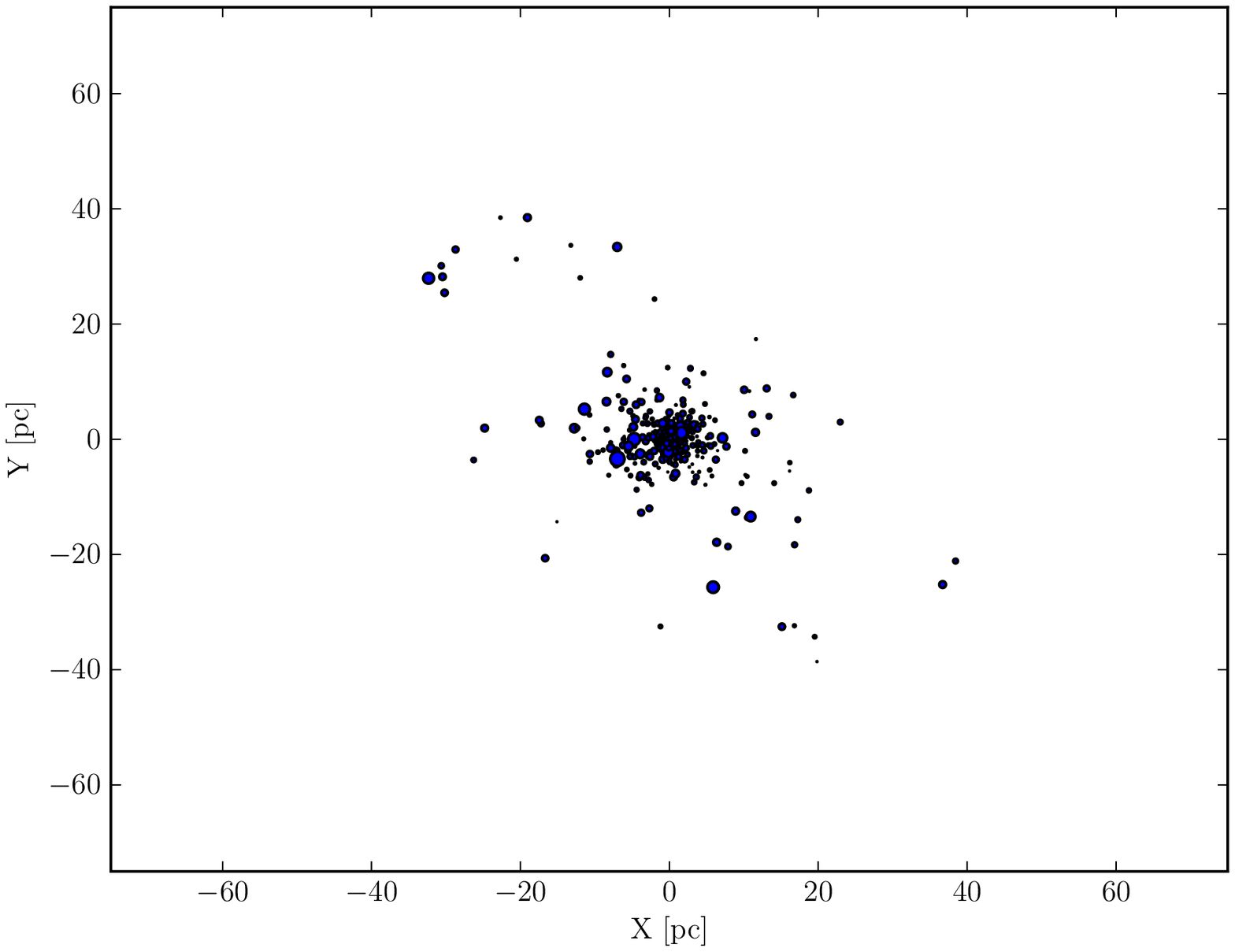}
\end{tabular}
\caption{As Figure \ref{appen1} but for our {\it{set1b}} model. The same rule for sizes was maintain for binary systems and their z-projection, although a slightly enhance on their size was produce to differentiate these points from single stars.}
\label{appen2}
\end{figure*}

\begin{figure*}
\centering
\begin{tabular}{cc}
\includegraphics[width=0.45\textwidth]{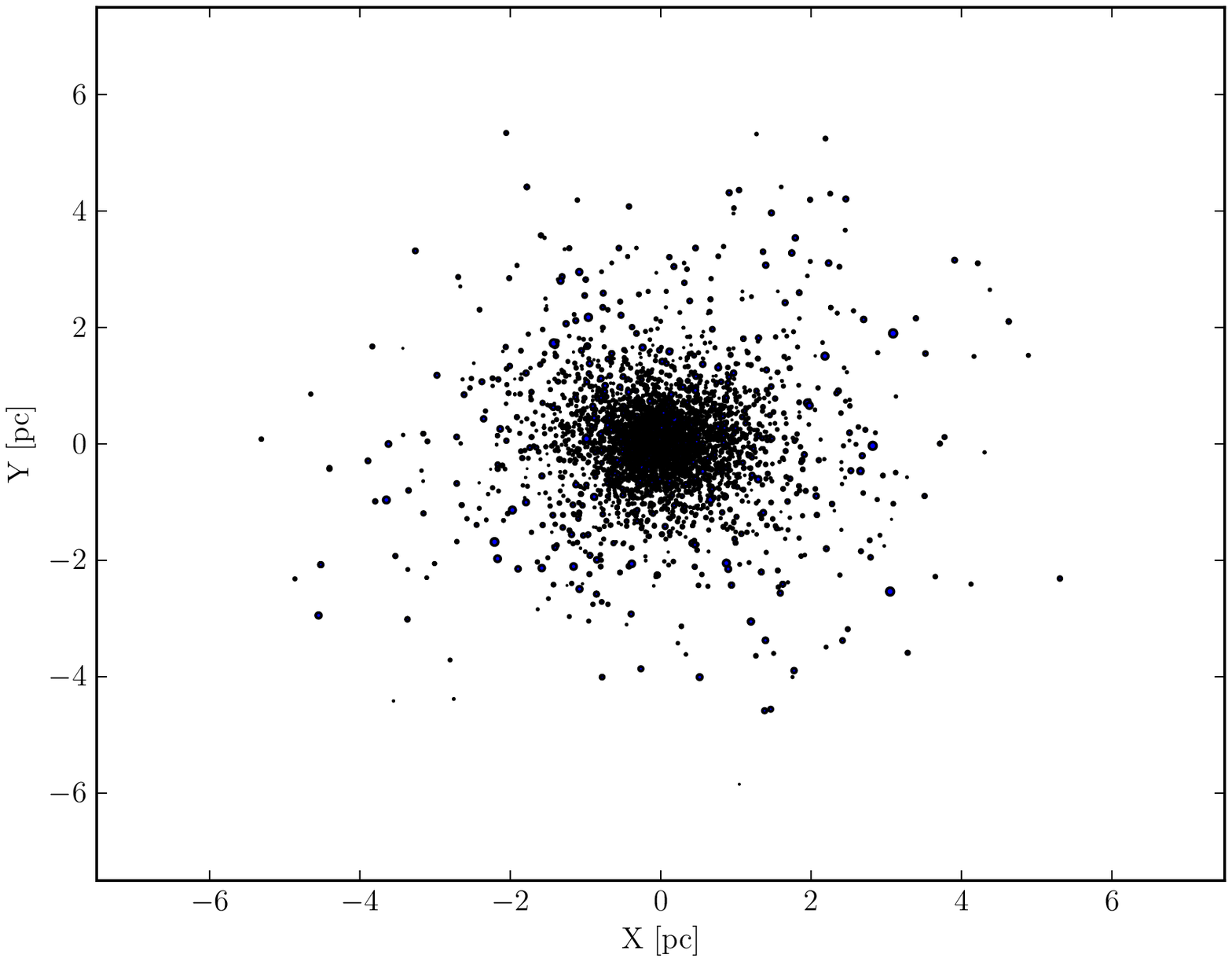} &
\includegraphics[width=0.45\textwidth]{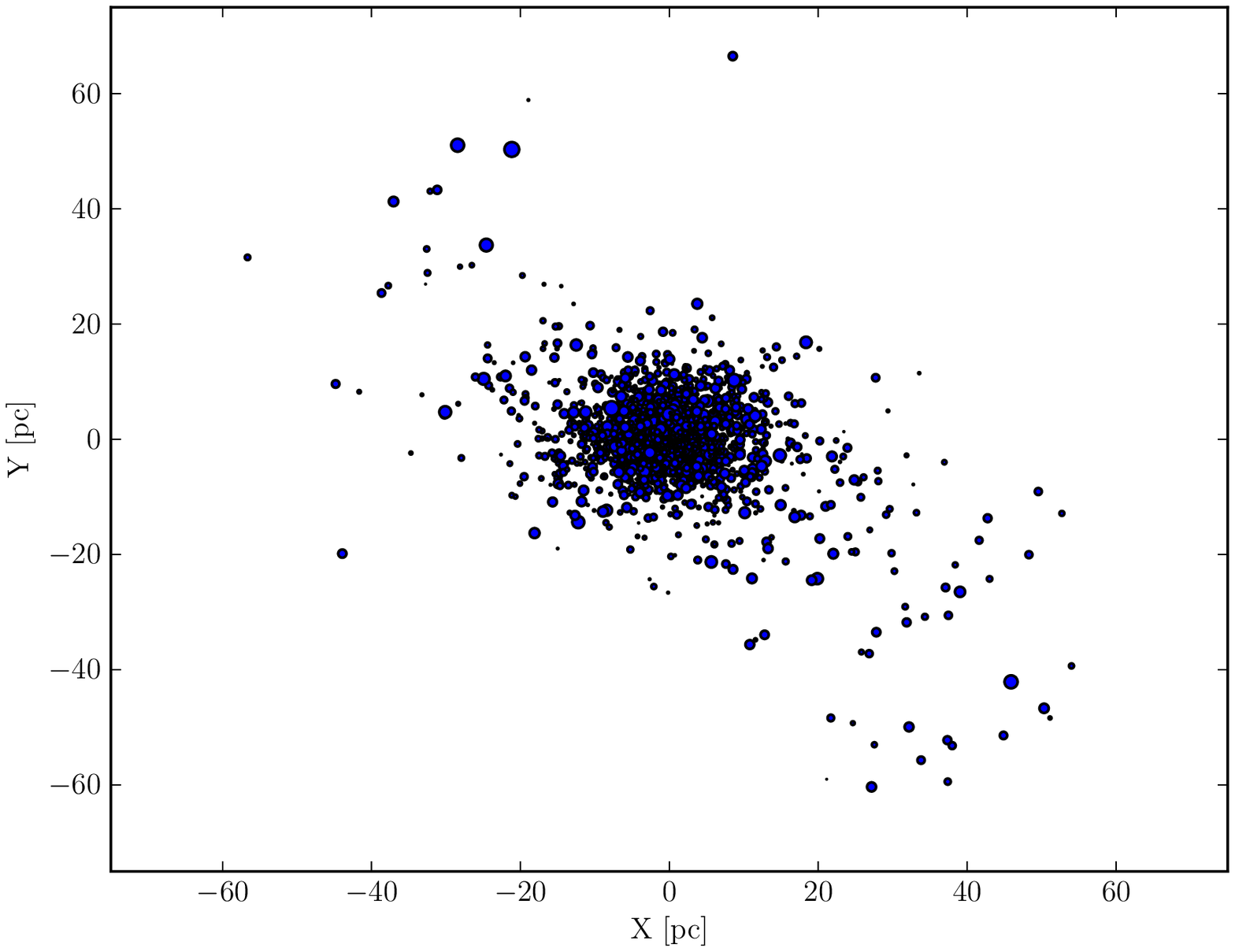} \\
\includegraphics[width=0.45\textwidth]{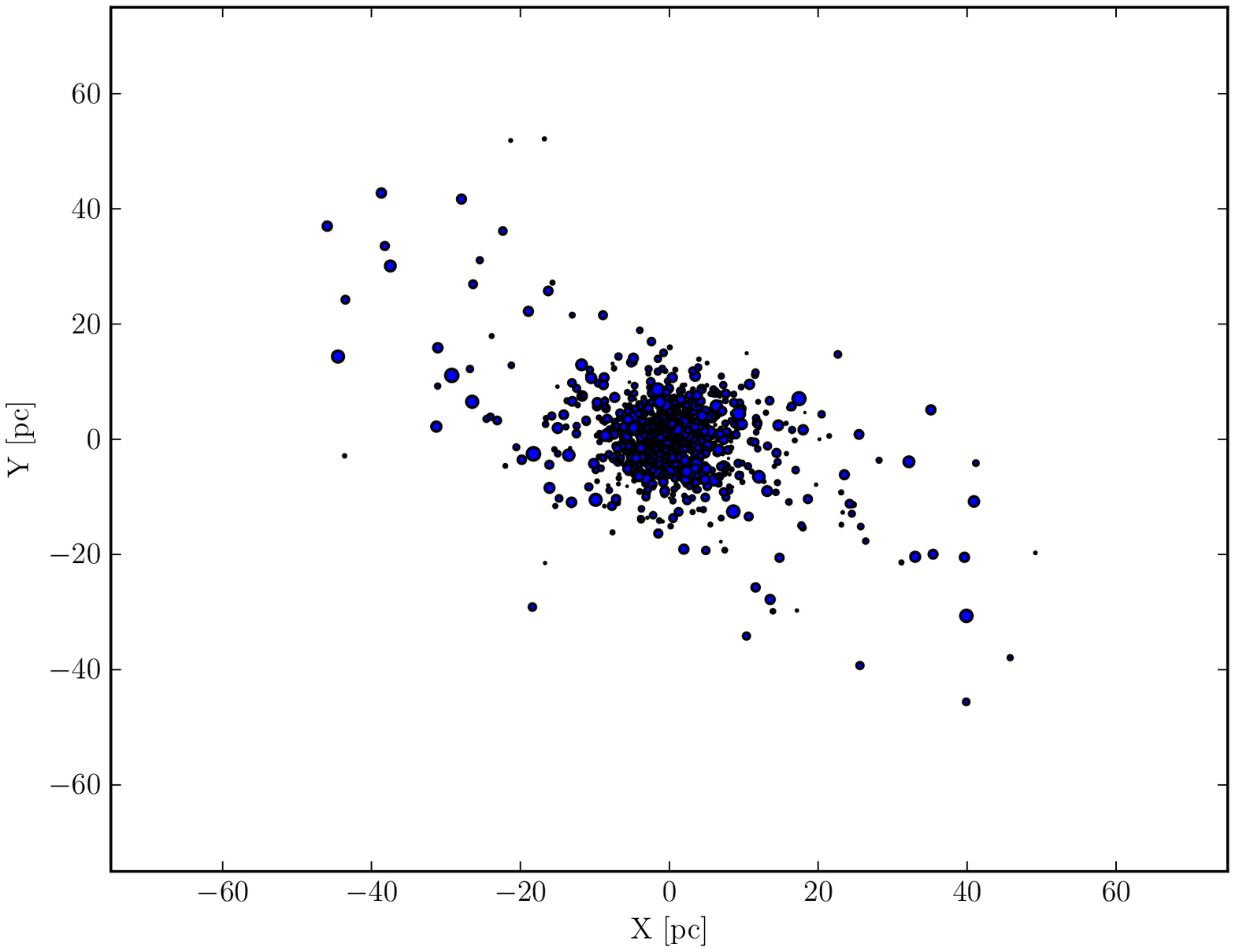} &
\includegraphics[width=0.45\textwidth]{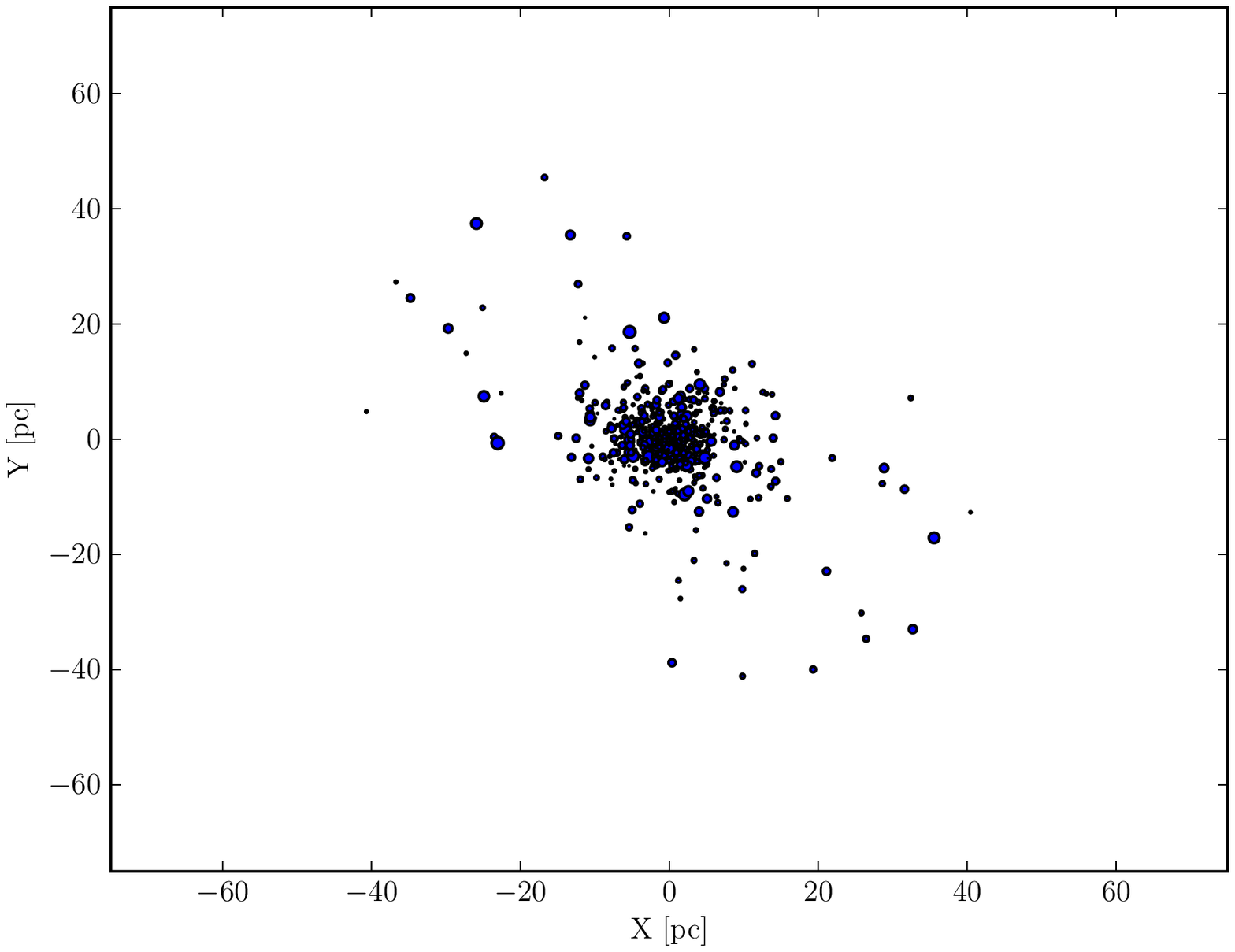}
\end{tabular}
\caption{As Figure \ref{appen2} but for our {\it{set1c}} model.}
\label{appen3}
\end{figure*}

% \bsp % ``This paper has been produced using the ...''

\label{lastpage}

\end{document}